\documentclass[fleqn,usenatbib]{mnras}

\usepackage{newtxtext,newtxmath}

\usepackage[T1]{fontenc}

\DeclareRobustCommand{\VAN}[3]{#2}
\let\VANthebibliography\thebibliography
\def\thebibliography{\DeclareRobustCommand{\VAN}[3]{##3}\VANthebibliography}

\usepackage{graphicx}	
\usepackage{amsmath}	

\usepackage{tikz} 
\usepackage{tabularx}
\usepackage{tabu}
\usepackage{makecell}
\usepackage{multicol,multirow}
\usepackage{xcolor}
\usepackage{ulem}
\usepackage{newtxtext,newtxmath}
\usepackage{threeparttable}

\newcolumntype{C}[1]{>{\centering\arraybackslash}p{#1}}
\defcitealias{pierce_2022}{P22}
\defcitealias{Desmons_2025}{Desmons et al. 2025}

\title[Triggering mechanisms of radio AGN sub-types]{The dependence of triggering mechanisms on radio AGN sub-types: the role of galaxy mergers}

\author[F. Barwell et al.]{F. Barwell,$^{1}$\thanks{E-mail: fbarwell1@sheffield.ac.uk}
C. N. Tadhunter,$^{1}$
J. C. S. Pierce,$^{2}$
A. E. Watkins,$^{2}$
Y. Gordon,$^{3}$
L. R. Holden,$^{2}$
L. Makrygianni,$^{4}$
\newauthor D. T. Mason,$^{1}$
A. J. Singleton,$^{1}$
R. J. Houghton,$^{1,5}$
S. A. J. McLaughlin,$^{1}$
C. Ramos Almeida$^{6,7}$
and J. Rom\'an$^{8,9}$\\ \\
$^{1}$Astrophysics Research Cluster, School of Mathematical and Physical Sciences, University of Sheffield, Sheffield S3 7RH, UK \\
$^{2}$Centre for Astrophysics Research, University of Hertfordshire College Lane, Hatfield AL10 9AB, UK \\
$^{3}$Department of Physics, University of Wisconsin-Madison, 1150 University Ave, Madison, WI 53706, USA \\
$^{4}$Department of Physics, Lancaster University, Lancaster, LA1 4YB, UK,\\
$^{5}$Astrophysics Research Institute, Liverpool John Moores University, IC2, Liverpool Science Park, 146 Brownlow Hill, Liverpool L3 5RF, UK\\
$^{6}$Instituto de Astrofísica de Canarias, Calle Vía Láctea, s/n, E-38205, La Laguna, Tenerife, Spain\\
$^{7}$Universidad de La Laguna, Calle Astrofísico Francisco Sánchez, E-30206, La Laguna, Tenerife, Spain\\
$^{8}$Departamento de Física de la Tierra y Astrofísica, Universidad Complutense de Madrid, E-28040 Madrid, Spain\\
$^{9}$Departamento de Física, Universidad de Córdoba, Campus de Rabanales, Edificio Albert Einstein, E-14071 Córdoba, Spain}

\date{Accepted XXX. Received YYY; in original form ZZZ}

\pubyear{\the\year{}}

\begin{document}
\label{firstpage}
\pagerange{\pageref{firstpage}--\pageref{lastpage}}
\maketitle

\begin{abstract}

\noindent{Powerful, radio-loud active galactic nuclei (AGN) are associated with one of the most important forms of AGN feedback, and understanding how they are triggered is key to properly incorporating them into models of galaxy evolution. Here, we present the results of a deep Isaac Newton Telescope/Wide Field Camera imaging survey which, when combined with Gemini/Gemini Multi-Object Spectrograph South images, gives a 98 per cent complete sample of 112 3CR radio galaxies with redshifts \textit{z} < 0.3, alongside a stellar mass matched control sample. Our results provide strong evidence for significant differences ($\sim$3$\sigma$) between the triggering mechanisms of the different sub-types of powerful radio AGN. The high-excitation radio galaxies (HERGs) show a high rate of morphological disturbance (62$^{+6}_{-7}$ per cent) -- an excess of $\sim$4$\sigma$ compared with the control sample -- consistent with them being predominantly triggered in galaxy mergers and interactions. In contrast, the low-excitation radio galaxies (LERGs) show a much lower rate of morphological disturbance (36$^{+7}_{-6}$ per cent), consistent with the control sample, and suggesting a different dominant triggering mechanism, such as the accretion of gas from the hot X-ray haloes of the host galaxies or galaxy clusters. We also demonstrate that, when considering the radio morphology, the FRII HERG sources preferentially reside in disturbed morphologies, a difference of $\sim$3$\sigma$ to the FRII LERG objects. This suggests that the FRII LERG sources do not solely represent a `switched-off' phase in the HERG lifecycle of the same parent galaxy population as the FRII HERGs.}

\end{abstract}




\section{Introduction}

Powerful, radio-loud active galactic nuclei (AGN) ($L_{1.4\mathrm{GHz}}>10^{24}$~W~Hz$^{-1}$) are a key subclass of the overall AGN population \citep[see][for a review]{Tadhunter_2016}. They show a strong preference for massive elliptical galaxies ($M_*>10^{11}M_\odot$), and are associated with one of the most important forms of AGN-induced feedback \citep[see][]{Best_2006, Hardcastle_Croston_2020, Harrison_CRA_2024}. Stellar mass growth is inhibited as their expanding radio jets and lobes prevent the hot X-ray emitting gas of the host galaxies and clusters from cooling to form stars. This affects the shape of the galaxy-luminosity and stellar mass functions at the high-luminosity/mass end \citep[e.g.][]{Bower_2006, Croton_2006, McNamara_Nulsen_2007}. However, despite being an important source of feedback, we do not fully understand how radio AGN are triggered as their galaxies evolve. 

As a first step to understanding the triggering of radio AGN, it is important to consider how they are classified at both optical and radio wavelengths. At optical wavelengths, they can be classified using the narrow-line region (NLR) ionisation conditions, as revealed by emission line ratios. Based on their excitation indices (EI), there are two main sub-populations of radio-loud AGN: high-excitation radio galaxies (HERGs) with EI $> 0.95$ and low-excitation radio galaxies (LERGs)\footnote{Alternatively, HERG and LERG sources are sometimes labelled as LEG (low-excitation galaxy) and HEG (high-excitation galaxy) objects \citep[e.g.][]{Buttiglione_2010}.} with EI $< 0.95$ \citep{Buttiglione_2010}. Radio AGN have also been classified as strong-line radio galaxies (SLRGs) and weak-line radio galaxies (WLRGs) according to whether their [OIII]$\lambda$5007 equivalent widths are larger or smaller than 10\AA, respectively \citep{Tadhunter_1998}. This scheme shows significant overlap with the HERG/LERG classification \citep[see discussion in][]{Tadhunter_2016}. More recently, other approaches have also been proposed to classify radio AGN as HERGs and LERGs based on various optical and infrared properties \citep[e.g.][]{Best_Heckman_2012, Mingo_2022, Best_2023}. 

As well as by their optical spectra, radio AGN can also be classified according to their radio morphology, most notably using the Fanaroff-Riley classification scheme \citep{Fanaroff_riley_1974}. FRI sources display edge-darkened morphologies in contrast to the edge-brightened FRII sources. There are also some objects which display a hybrid FRI/FRII morphology \citep{GoKri_2000}. FRI sources are almost invariably associated with LERGs and HERGs with FRII sources. However, the mapping between the optical and radio classifications is not perfect, since a significant subset ($\sim$20-40 per cent) of FRII sources have been classified as LERGs \citep[e.g.][]{Buttiglione_2010, Tadhunter_2016}.

Understanding how the different sub-types of radio AGN are triggered as their host galaxies evolve is key to accurately incorporate their feedback effects into models of galaxy evolution. Previous studies have suggested that some of the differences between the sub-types of radio AGN are due to differing rates of accretion onto the supermassive black hole (SMBH), caused by a change in the dominant triggering mechanism \citep{Hardcastle_2006, Heckman_Best_2014}. Radiatively-efficient and -inefficient AGN represent two distinct modes of accretion onto the SMBH.

In radiatively-efficient AGN (e.g. HERGs, SLRGs, quasars), material is accreted via an optically thick, geometrically thin accretion disk surrounding the SMBH \citep[e.g.][]{Shakura_Sunyaev_1973} at a high Eddington rate \citep[$>1$ per cent;][]{Best_Heckman_2012, Mingo_2014, Arnaudova_2025}. Tidal torques associated with galaxy mergers have been suggested as a viable means of supplying a sufficient inflow of cold gas, capable of maintaining the required high Eddington accretion rate to trigger and sustain the AGN activity \citep[e.g.][]{Barnes_Hernquist_1996, Springel_2005, Gabor_2016}. Indeed, deep ground-based imaging studies of powerful radio AGN ($L_{\mathrm{1.4GHz}}>10^{25}$ W Hz$^{-1}$) have found morphological evidence for a high rate of galaxy mergers or interactions \citep{Heckman_1986, Smith_Heckman_1989_colours_1, Smith_Heckman_1989_colours_II, ramos_almeida_2011, ramos_almeida_2012, pierce_2022}. For example, \cite{pierce_2022} \citepalias[hereafter][]{pierce_2022} found $66^{+7}_{-8}$ per cent of the 3CR HERGs in their sample displayed clear signs of morphological disturbance, exhibiting a $\sim$5$\sigma$ excess compared with a stellar mass and redshift matched control sample. Moreover, characterisation of the environments of the radiatively-efficient AGN have revealed they are preferentially hosted in group-like environments, where the conditions are more favourable for galaxy mergers or interactions \citep{Ramos_almeida_2013}. 

In contrast, radiatively-inefficient AGN (e.g. LERGs, WLRGs) accrete material at a much lower Eddington rate \citep[$<1$ per cent;][]{Best_Heckman_2012, Mingo_2014, Arnaudova_2025}, via an optically thin, geometrically thick accretion disk \citep[e.g.][]{Narayan_Yi_1994}. The 3CR LERGs from \citetalias{pierce_2022} displayed a much lower proportion of galaxy mergers and interactions, with only $37^{+9}_{-8}$ per cent showing clear tidal features -- consistent with that of a stellar mass and redshift matched control sample. Moreover, characterisation of the LERG environments at both optical \citep{Ramos_almeida_2013} and X-ray wavelengths \citep{Ineson_2013, Ineson_2015} revealed that these objects favour cluster-like environments. Due to the high relative galaxy velocities, dense cluster-like environments are not favourable for galaxy mergers \citep[e.g.][]{Popesso_Biviano_2006}. The dominant triggering mechanisms of these objects have therefore been suggested to be associated with the abundance of hot gas in their environments, via direct hot gas accretion \citep[e.g.][]{Allen_2006, Hardcastle_2007}, cooling flows \citep[e.g.][]{Tadhunter_1989, Baum_1992} or the chaotic accretion of cold condensing gas \citep[e.g.][]{Gaspari_2013, Gaspari_2015}. 

These results suggest that there are differences between the dominant triggering mechanisms between the different sub-types of radio AGN. However, the levels of statistical significance for this difference have so far been relatively low. For example, \citetalias{pierce_2022} found that the difference between the proportions of the disturbed HERGs and the disturbed LERGs was only significant at the $\sim$2.4$\sigma$ level. This is largely due to small sample sizes, particularly for the radiatively-inefficient radio AGN. In addition, any separate triggering dependence with the radio morphologies has not been previously investigated due to small sample sizes of FRI sources. 

In this work, the 3CR sample from \citetalias{pierce_2022}, which covered redshifts $0.05<z<0.3$, is expanded to include all objects with lower redshifts ($z<0.05$). This greatly increases the number of objects with LERG classifications (from 30 to 58 objects) and FRI radio morphologies (from 3 to 23 objects). In addition, we include 10 3CR/2Jy\footnote{Objects in the 2Jy sample were selected to have a flux density $S_\mathrm{2.7 GHz}>2$Jy and declinations $\delta<+10^{\circ}$ \citep{Wall_Peacock_1985}, whereas sources in the 3CR sample were selected to have a flux density $S_\mathrm{178 MHz}>9$Jy and declinations $\delta>-5^{\circ}$ \citep{bennett_1962_2, bennett_1962_1}. There is some overlap between the two catalogues. We refer to such overlapping objects in our sample as 3CR/2Jy sources.} objects from \cite{ramos_almeida_2011}, resulting in a 98 per cent complete sample of 112 powerful 3CR radio AGN -- which represents the most luminous radio AGN population ($L_{1.4\mathrm{GHz}}>10^{24}$~W~Hz$^{-1}$) -- at redshifts $z<0.3$. We performed a morphological analysis of this sample, alongside a stellar mass matched control sample of 307 galaxies selected from within the same image fields. The inclusion of both radiatively-efficient and -inefficient radio AGN within the sample allows the determination, with robust statistics, of whether there are truly differences in the dominant triggering and fuelling mechanisms of these objects. 

The paper is structured as follows. In Section~\ref{sample_obs_reduction} we discuss the sample selection, observations and data reduction. In Section~\ref{control_match+classification} we outline the selection of the control sample and the methodology used to obtain our morphological classifications. Our analysis and results are detailed in Section~\ref{analysis+results}. Finally, a discussion of the results is presented in Section~\ref{discussion}, and conclusions in Section~\ref{conclusions}. A cosmology with $H_0=73$km s$^{-1}$ Mpc$^{-1}$, $\Omega_\mathrm{m}=0.27$ and $\Omega_\Lambda=0.73$, is assumed for consistency with previous work. 

\section{Sample, observations and reduction}
\label{sample_obs_reduction}

\subsection{3CR sample}
\label{sample_info}

Our sample of radio AGN consists of all the 3CR radio AGN included in the \cite{Buttiglione_2009, Buttiglione_2010, Buttiglione_2011} sample -- originally selected from \cite{Spinrad_1985} -- with redshifts less than $z<0.3$, with a few exceptions. We do not include 3C 273 from \cite{Buttiglione_2009, Buttiglione_2010, Buttiglione_2011} due to its nature as the optically brightest quasar, making it difficult to perform a detailed morphological analysis of the host galaxy. The galaxy 3C 452 could not be included due to the image being compromised at the galaxy location by a nearby saturated star. However, we do include 3C 405 (Cygnus A), which was not included in the \cite{Buttiglione_2009, Buttiglione_2010, Buttiglione_2011} sample, but is present in the \cite{Spinrad_1985} catalogue. 

Our previous work in \citetalias{pierce_2022}, used observations from the Isaac Newton Telescope/Wide Field Camera (INT/WFC) to study the host galaxies of 72 objects in the higher redshift part of this sample with redshifts $0.05 < z < 0.3$, excluding the 10 objects that are also in the 2Jy sample of \cite{ramos_almeida_2011} and already had imaging observations with the Gemini Multi-Object Spectrograph South (GMOS-S) on the Gemini South telescope. In this work, we extend their sample to cover all the remaining 3CR sources redshifts less than $z<0.05$ -- a total of 30 sources, and the 10 3CR/2Jy objects that were observed using Gemini South.

Although the observations of the 10 3CR/2Jy sources were conducted by GMOS-S under better seeing conditions, the observations had a similar surface brightness depth, and the results from the images of these sources presented in the online classification interface (see Section~\ref{zooniverse}) were largely consistent with the WFC-imaged sources (refer to Section~\ref{merger_detection_rate}). Therefore, it is not expected that any major biases were introduced through the inclusion of these sources. The addition of these objects gives a 98 per cent complete sample of 112 3CR sources with redshifts less than $z<0.3$. 

\begin{figure}
    \centering
    \includegraphics[width=\linewidth]{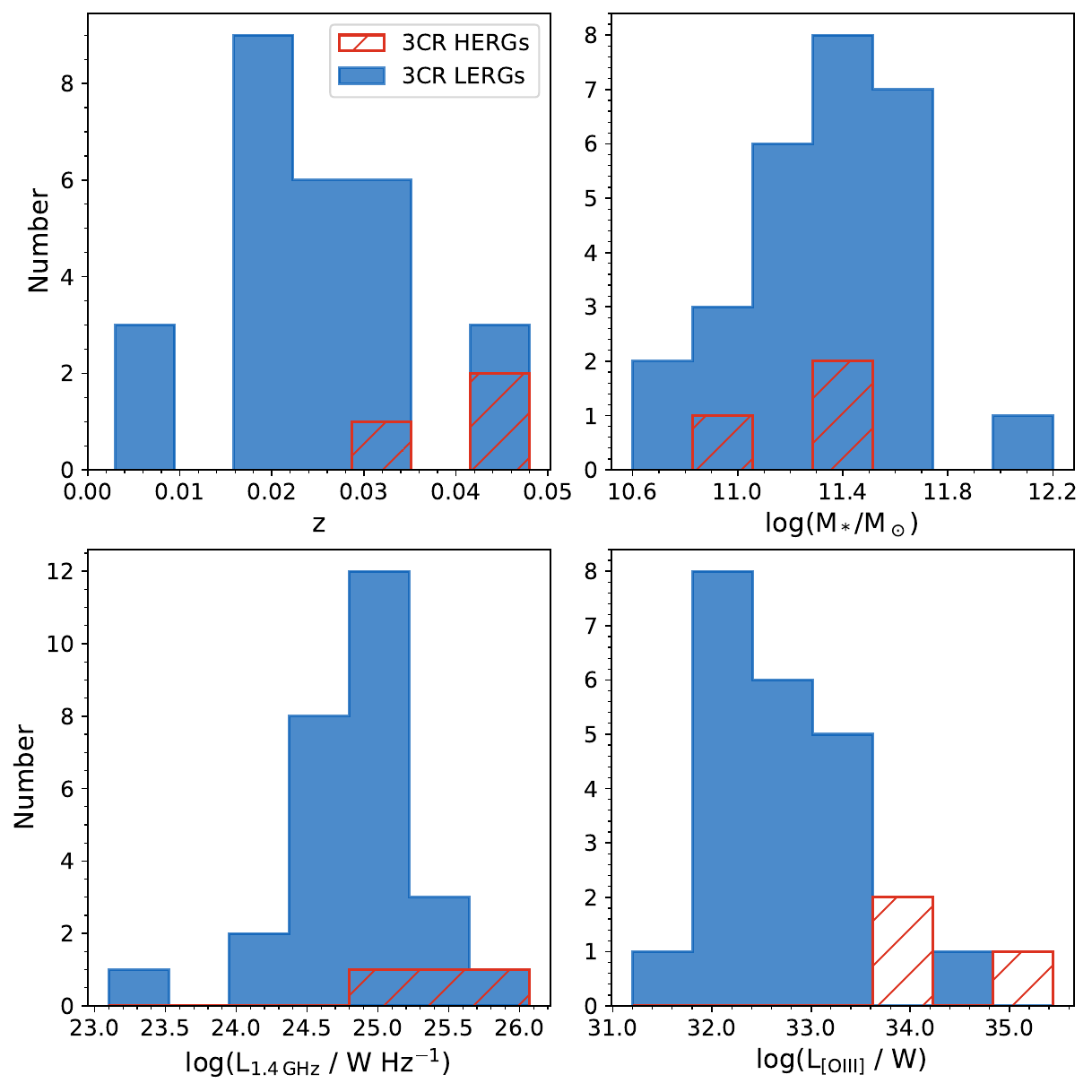}
    \caption{Distributions of redshift, stellar mass, 1.4 GHz radio luminosity and [OIII]$\lambda$5007 emission-line luminosity for the HERGs (red) and the LERGs (blue) in the low-redshift ($z<0.05$) 3CR sample. Sources with no stellar mass estimates were not included in the corresponding plot. Those which only had upper limits on their [OIII]$\lambda$5007 emission-line luminosity were also not considered in the distribution.}
    \label{fig:low_z_3CR_distributions}
\end{figure}

\begin{figure}
    \centering
    \includegraphics[width=\linewidth]{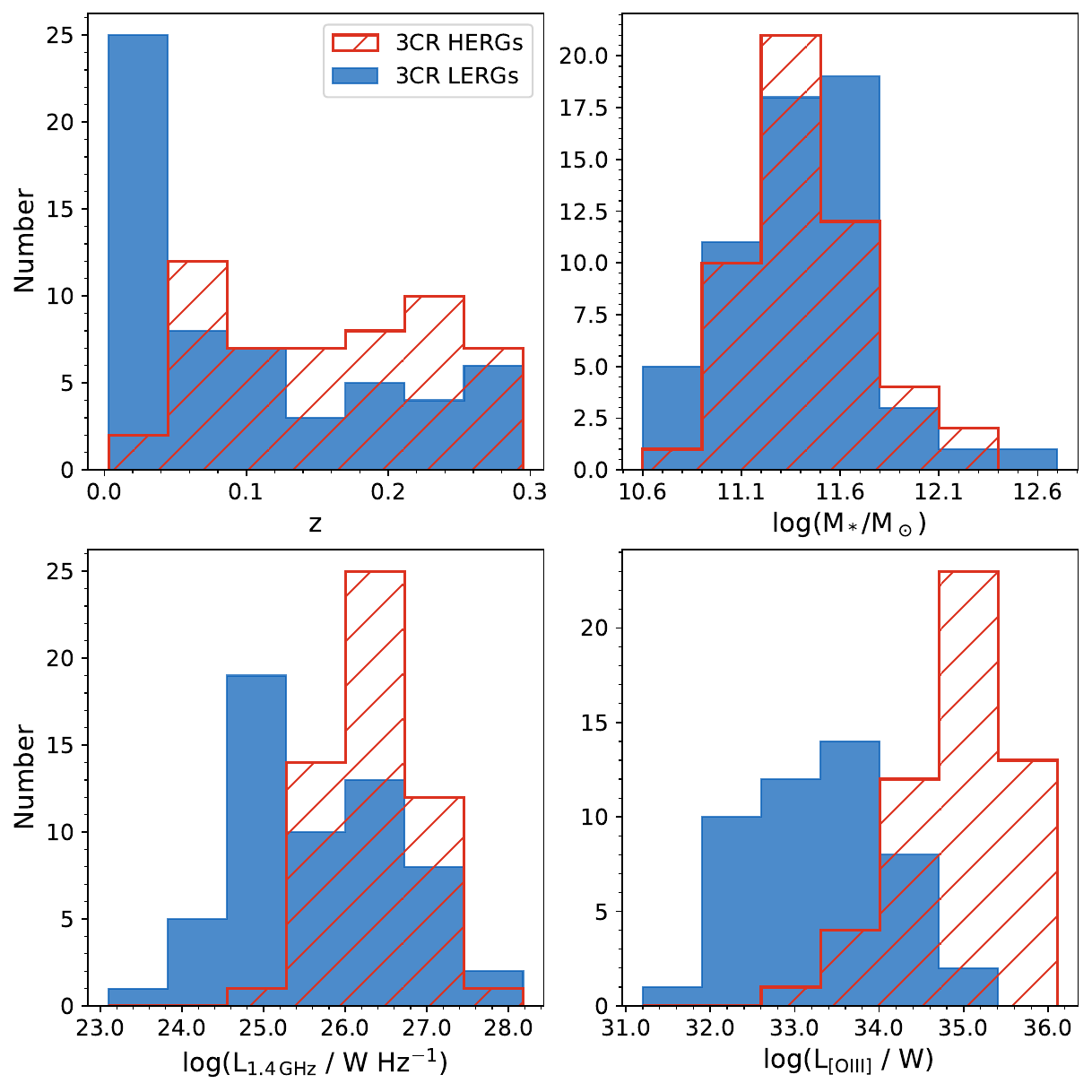}
    \caption{As in Figure~\ref{fig:low_z_3CR_distributions}, but for the full 3CR sample ($z<0.3$) considered in this work.}
    \label{fig:full_3CR_distributions}
\end{figure}

Basic information about the 30 low-\textit{z} sources ($z<0.05$) is listed in Table~\ref{tab:sample_info_full}, while information regarding the other 72 3CRs and the 10 3CR/2Jy objects in the sample are included in the supplementary material in \citetalias{pierce_2022} and in  \cite{ramos_almeida_2011}, respectively. Redshift, stellar mass, 1.4 GHz radio luminosity and [OIII]$\lambda$5007 emission-line luminosity distributions for the HERGs and LERGs in the low-\textit{z} and full 3CR sample are presented in Fig.~\ref{fig:low_z_3CR_distributions} and Fig.~\ref{fig:full_3CR_distributions}, respectively, whereas the distributions for the FRIs and FRIIs in the low-\textit{z} and full 3CR sample are presented in Appendix~\ref{fr1_fr2_dists}. The measured [OIII]$\lambda5007$ emission-line luminosities for the full sample were mainly obtained from \cite{Buttiglione_2009, Buttiglione_2010, Buttiglione_2011}, with a few exceptions for the higher redshift portion of the sample, discussed in \citetalias{pierce_2022}. In the low-$z$ portion of the sample, no [OIII]$\lambda5007$ line was detected in 3C 129.1, therefore, the H$\alpha$ emission-line flux was used to provide an upper limit constraint on the [OIII]$\lambda5007$ luminosity, assuming the noise levels were the same in both spectral regions. 

Of the 112 objects in the full sample, there are 26 sources that have quasar-like luminosities according to the [OIII]$\lambda$5007 emission-line criteria for Type 2 quasars given by \cite{Zakamska_2003} (L$_{\mathrm{[OIII]}}\geq10^{35}$~W). The 1.4 GHz radio luminosities were estimated from the 178 MHz radio luminosities given in \cite{Buttiglione_2009, Buttiglione_2010, Buttiglione_2011}, assuming a radio spectral index of $\alpha=-0.7$ (for $S_\nu\propto\nu^{+\alpha}$), and were converted to the cosmology used throughout this work. 

Our optical classifications (HERGs and LERGs) were mostly obtained from the classifications using the EI method from \cite{Buttiglione_2010, Buttiglione_2011}, with a some exceptions for the higher redshift portion of our sample discussed in \citetalias{pierce_2022}. For the low-\textit{z} sample, there were some objects which lacked detections of key spectral lines for determination of the EI (3C 75N, 3C 76.1, 3C 83.1, 3C 129, 3C 129.1, 3C 318.1, 3C 386 and 3C 402). We consider these as more extreme low-excitation sources, and as such include them under the LERG classification. Our radio classifications are mainly the same as those in \cite{Buttiglione_2010,Buttiglione_2011}; however, the radio classifications in these papers are incomplete. Therefore, we have completed and updated them using radio maps available in the literature (see Appendix~\ref{classification_results}).

Based on their optical classification, our sample comprises 53 HERGs, 58 LERGs and one star-forming galaxy. In terms of their radio classifications, the sample consists of 82 FRIIs\footnote{Our FRII classification includes compact steep spectrum (CSS) sources \citep[see][]{O_Dea_Saikia_2021}, which are compact (D < 15 kpc), but have morphologies similar to FRII sources.}, 23 FRIs, four hybrid FRI/FRII sources, two core-halo/jet morphologies and one `double-double' source. In our subsequent analysis, we consider the sources with hybrid FRI/FRII morphologies and the `double-double' morphology as FRII sources. 

\begin{table*}
	\centering
    \caption{Host-galaxy/AGN properties and observation details for the 30 low-redshift ($z<0.05$) 3CR targets in the sample. Column key: (1) 3CR catalogue name; (2) redshift \protect\citep[from][]{Spinrad_1985}; (3) 1.4 GHz radio luminosity \protect\citep[adapted from][assuming a spectral index of $\alpha=-0.7$]{Buttiglione_2009, Buttiglione_2010, Buttiglione_2011}; (4) [OIII]$\lambda5007$ emission line luminosities \protect\citep[from][]{Buttiglione_2009, Buttiglione_2010, Buttiglione_2011}; (5) MPA-JHU equivalent stellar mass estimates (see Section \ref{fig:stellar_mass}); (6) optical classification (HERG/LERG), with quasar-like AGN indicated (L$_{\mathrm{[OIII]}}\geq 10^{35}$ W); (7) radio classification (1: FRI; 2: FRII; CSS: Compact Steep Spectrum; DD: double-double); (8) observation date; (9) exposure time and (10) $r$-band atmospheric seeing measurements.}
    \label{tab:sample_info_full}
    \begin{threeparttable}
    \begin{tabular}{lccccccccc} %
		\hline
            \makecell{(1)\\Name\\{}} & \makecell{(2)\\ \textit{z}\\{}} & \makecell{(3)\\ log$(L_{1.4~\textrm{GHz}}/$  \\  $\textrm{W\,Hz}^{-1})$} & \makecell{(4) \\ log$(L_{\rm [OIII]}/$ W)\\{}} & \makecell{(5) \\ log($M_{*}$/$M_{\odot}$) \\{}} & \makecell{(6) \\ Optical class\\{}} & \makecell{(7) \\ Radio class\\{}} & \makecell{(8) \\ Date\\{}}  &  \makecell{(9) \\ Exp. time \\(s)} & \makecell{(10) \\ Seeing \small{FWHM}\\ (arcsec)} \\ 

		\hline
		      3C 29 & 0.045 & 24.19 & 33.07 & 11.4 & LERG & 1 & 2022-07-29 & 4 $\times$ 700 & {1.86}\\
		      3C 31 & 0.017 &  24.36 & 32.44 & 11.4 & LERG & 1 & 2022-07-30 & 4 $\times$ 700 & {1.31}\\
		      3C 40 & 0.018 & 24.64 & 32.20 & 11.4 & LERG & 1 & 2022-07-30 & 4 $\times$ 700 & {1.35}\\
            3C 66B & 0.021 & 24.75 & 33.03 & 11.4 & LERG & 1 & 2022-07-31 & 6 $\times$ 700 & {1.29}\\
            3C 75N & 0.023 & 24.84 & < 32.90 & 11.5 & LERG & 1 & 2022-07-29 & 4 $\times$ 700 & {2.10}\\
            3C 76.1 & 0.032 &  24.81 & < 32.83 & 10.6 & LERG & 1& 2022-07-30 & 4 $\times$ 700 & {1.64}\\
            3C 78 & 0.028 & 24.86 & 32.39 & 11.7 & LERG & 1 & 2022-07-31 & 4 $\times$ 700 & {1.57}\\
            3C 83.1 & 0.026 & 24.92 & < 32.48 & 11.7 & LERG & 1 & 2022-08-22 & 4 $\times$ 700 & {1.58}\\
            3C 84 & 0.017 & 24.97 & 34.58 & 11.6 & LERG & 1 & 2022-08-22 & 4 $\times$ 700 & {1.41}\\
            3C 88 & 0.030 & 24.84 & 33.12 & 11.1 & LERG & 2 & 2022-08-22 & 4 $\times$ 700 & {1.60}\\
            3C 98 & 0.032 & 25.34 & 33.98 & 11.0 & HERG & 2 & 2022-12-18 & 4 $\times$ 700 & {1.42}\\
            3C 111 & 0.048 & 25.89 & 35.42 & 11.4 & HERG/Q & 2 & 2022-12-18 & 4 $\times$ 700 & {1.45}\\
            3C 129 & 0.021 & 25.00 & 32.83 & 11.2 & LERG & 1 & 2022-12-18 & 4 $\times$ 700 & {1.37}\\
            3C 129.1 & 0.022 & 24.41 & < 32.81 & 12.2 & LERG & 1 & 2022-12-18 & 4 $\times$ 700 & {1.48}\\
            3C 264 & 0.020 & 24.78 & 32.18 & 11.3 & LERG & 1 & 2022-12-18 & 4 $\times$ 700 & {1.82}\\
            3C 270 & 0.007 & 24.14 & 31.96 & 11.3 & LERG & 1 & 2022-12-18 & 4 $\times$ 700 & {1.80}\\
            3C 272.1 & 0.003 & 23.07 & 31.18 & 11.1 & LERG & 1 & 2022-12-18 & 4 $\times$ 700 & {1.71}\\
            3C 274 & 0.004 & 24.98 & 31.97 & 11.5 & LERG & 1 & 2022-12-18 & 4 $\times$ 700 & {1.69}\\
            3C 293 & 0.045 & 25.12 & 32.78 & 11.3 & LERG & DD\tnote{*} & 2022-12-19 & 4 $\times$ 700 & {1.37}\\
            3C 296 & 0.025 & 24.57 & 32.76 & 11.7 & LERG & 1 & 2022-07-30 & 4 $\times$ 700 & {1.41}\\
            3C 305 & 0.042 & 25.14 & 34.01 &11.3 & HERG & CSS & 2022-07-29 & 4 $\times$ 700 & {1.53}\\
            3C 317 & 0.035 & 25.47 & 33.33 & 11.7 & LERG & core-halo & 2022-07-30 & 4 $\times$ 700 & {1.27}\\
            3C 318.1 & 0.044 & 25.07 & 32.34 & 11.3 & LERG & 2 & 2022-07-29 & 4 $\times$ 700 & {1.77}\\
            3C 338 & 0.031 & 25.34 & 32.55 & 11.7 & LERG & 1 & 2022-07-30 & 4 $\times$ 700 & {1.17}\\
            3C 353 & 0.030 & 26.04 &  33.12 & 11.0 & LERG & 2 & 2022-07-31 & 4 $\times$ 700 & {1.16}\\
            3C 386 & 0.017 & 24.53 & < 33.18 & 10.9 & LERG & 2 & 2022-07-29 & 4 $\times$ 700 & {2.10}\\
            3C 402 & 0.025 & 24.46 & < 32.40 & 11.0 & LERG & 1 & 2022-07-31 & 6 $\times$ 700 & {1.14}\\
            3C 442 & 0.026 & 24.74 & 32.19 & 11.2 & LERG & 2 & 2022-07-29 & 4 $\times$ 700 & {2.00}\\
            3C 449 & 0.017 & 24.22 & 32.17 & 11.0 & LERG & 1 & 2022-07-30 & 4 $\times$ 700 & {1.16}\\
            3C 465 & 0.030 & 25.24 & 32.79 & 11.6 & LERG & 1 & 2022-07-31 & 8 $\times$ 700 & {1.30}\\
	\hline
	\end{tabular}
    \begin{tablenotes}
    \item[*] 3C 293 has a 'double-double' (DD) radio morphology, with an inner and outer set of radio lobes \citep{Akujor_1996}. In our subsequent analysis, we consider it as a FRII source. 
    \end{tablenotes}
    \end{threeparttable}
\end{table*}

 \subsection{Observations}
 \label{observations}

The deep optical imaging data were obtained with the Wide-Field Camera (WFC), an optical mosaic camera mounted on the \mbox{2.54-m} Isaac Newton Telescope (INT) at the Roque de los Muchachos observatory, La Palma. The WFC consists of 4 thinned EEV42 4k$\times$2k CCDs, giving a total field of view of 34 $\times$ 34 arcmin$^{2}$, with a pixel scale of 0.333 arcsec pixel$^{-1}$. 

All the images of the low-\textit{z} 3CR sample were taken using the WFC Sloan \textit{r}-band filter ($\lambda_\mathrm{eff}=6240$\AA, $\Delta\lambda = 1347$\AA), consistent with the vast majority of observations of the 3CR sample in \citetalias{pierce_2022} and the radio-loud 2Jy sample in \cite{ramos_almeida_2011}.

The observations of the low-\textit{z} 3CR sample were taken during three separate observing runs in July, August and December 2022. The observation dates, and seeing estimates as determined from the average of the full width at half maximum (FWHM) measurements of foreground stars in the final coadd images (after the reduction outlined in Section~\ref{image_reduction}) for each target are provided in Table~\ref{tab:sample_info_full}. The low-\textit{z} 3CR sample had a median seeing of 1.46 arcsec, with a standard deviation of 0.27 arcsec. 

Most of the sample were observed for a total of 2800~s, using a 4~$\times$~700~s dithered exposures to avoid the target galaxy saturating and improve overall image quality. A few objects required additional exposures to increase the signal-to-noise in poorer observing conditions. The exposure times for each of the low-\textit{z} objects are given in Table~\ref{tab:sample_info_full}. A large integration time was necessary to enable the detection of any low surface brightness tidal features, and the observations reach a consistent limiting surface brightness depth of 27 mag arcsec$^{-2}$ \citepalias[measurements from][]{pierce_2022}. To fill the gaps between the four WFC CCDs, a square dithering offset pattern (30 arcsec in each direction) was used. 

\subsection{Image reduction}
\label{image_reduction}

The reduction, calibration and co-addition of the WFC target images was conducted using the automated reduction software, \textsc{theli} \citep{Schirmer_2013}. The biases and flat-field frames corresponding to each target image were median combined into a master bias and master flat-field frame. The target images were then bias corrected and flat-fielded by the subtraction and division of the master bias frame and master flat-field frame, respectively. 

From each image, an object catalogue was extracted using \textsc{SExtractor} \citep{Bertin_1996}. The \textsc{scamp} pipeline was then used within \textsc{theli} to cross-match this catalogue with the GAIA DR3 catalogue \citep{GAIA_mission_2016, GAIA_DR3_2023} and determine the astrometric solutions. 

Any remaining variations in the sky background of the calibrated images were then removed by subtracting a model of the sky produced by \textsc{theli}. As part of this process, objects that had a minimum of 5 connected pixels at $1.5\sigma$ above the background level were detected by \textsc{SExtractor} and removed. The best results were produced when the sky model was convolved with a Gaussian kernel with a FWHM of 150 pixels. 

This method worked well for a large proportion of the sample. However, for 5 of the targets (3C 449, 3C 40, 3C 270, 3C 272.1 and 3C 274), the sky model produced by \textsc{theli} caused areas of over-subtraction close to the targets. As our work focuses on the search for low surface brightness tidal features, it was imperative that the sky subtraction, especially in areas directly around the target, was accurate. Therefore, we used a separate method to model and subtract the sky for these 5 objects, as follows.

\textsc{theli} has the advantage that the output images of each processing stage are saved, thus the images prior to sky subtraction could be accessed. Sources in the target fields were first masked out using \textsc{NoiseChisel} \citep{Akhlaghi_2015_noisechisel}. The unmasked pixels were then modelled using a Legendre polynomial (orders were chosen on a case-by-case basis depending on the background's apparent complexity) and then subsequently subtracted from the images to correct for the sky background. The background modelling and masking process was iterated to improve the accuracy of the results. After subtracting the initial background model, NoiseChisel was re-run to generate updated masks from the sky-subtracted images. These revised masks were then used to re-derive the sky backgrounds from the original images. The objective was to minimize contamination from background flux in the masks, while maximising the inclusion of low-surface-brightness galaxy wings. However, it was observed that the second iteration of the sky models did not result in significant changes compared to the first. The sky-background-subtracted images were then fed back into \textsc{theli} for the final coaddition to keep the reduction as consistent as possible with the other images.

All photometric zeropoints were also determined in \textsc{theli}. The derived instrumental magnitudes for stars in the target images were compared with their catalogued magnitudes from the ATLAS-REFCAT2 \citep{Tonry_2018}. As the calibration stars were observed under the same conditions and in the same fields as the target sources, any photometric variability was automatically corrected for; this removed the need for separate observations of standard stars. 

\section{Control matching and classification methodology}
\label{control_match+classification}

\subsection{Stellar mass determination}

The control galaxies used for comparison with the 3CR sample were selected from the value-added catalogue from the Max Planck Institute for Astrophysics and the Johns Hopkins University (the MPA-JHU value-added catalogue) \citep{Kauffmann_2003, Brinchmann_2004, Tremonti_2004, Salim_2007}, based on spectral analysis of SDSS DR7\footnote{Available at: \url{https://wwwmpa.mpa-garching.mpg.de/SDSS/DR7/}.} \citep{York_2000, Abazajian_2009}. In order to select suitable control galaxy candidates, estimates of the stellar mass for the low redshift 3CR targets were required. Our methods followed those of \citetalias{pierce_2022}, to which we refer the reader for full details.

As the MPA-JHU catalogue did not contain stellar mass estimates for all the 3CR sources, we estimated their stellar masses by converting the Two Micron All Sky Survey \citep[2MASS;][]{Skrutskie_2006} $K_s$-band luminosities, using the colour-dependent mass-to-light ratio formula from \cite{Bell_2003}. We assumed a $B-V$ colour of 0.95, as expected for a typical elliptical galaxy at redshift zero \citep{Smith_Heckman_1989_colours_II}, and a Kroupa initial mass function \citep{Kroupa_2001}. 

We used the $K_s$-band magnitudes from the 2MASS Extended Source Catalogue \citep{Jarrett_2000} to determine the galaxy luminosities (XSC magnitudes, hereafter). The XSC magnitudes were available for 29 of the 30 3CR targets. For the remaining target (3C~402), the magnitude from the 2MASS Point Source Catalogue (PSC magnitudes, hereafter) was used instead \citep{Skrutskie_2006}. To correct for any potential missed flux, we estimated a correction derived from the average difference between the XSC and PSC magnitudes for the all the galaxies in the MPA-JHU catalogue (where available) within the redshift range of the 3CR sample ($z<0.05$). A median magnitude difference of $K_s^{PSC} - K_s^{XSC} = 1.101 $ was subtracted from the PSC magnitude to convert it to an estimated XSC magnitude. In addition, all the $K_s$-band magnitudes were corrected for extragalactic extinction, and K-corrected using the formula from \cite{Bell_2003}.

No correction was made for possible AGN contamination, since at $K_s$-band wavelengths (2MASS: 2.159$\mu$m) contributions from Type 2 AGN have been found to be generally small \citep[][]{Ramirez_2014_a, Ramirez_2014_b}. Moreover, Type 1 AGN only comprise 3 per cent of the low-$z$ 3CR sample. Therefore, a major impact on the main results is not expected.

We converted the $K_s$-band stellar mass estimates to equivalent MPA-JHU values using the prescription described in \citetalias{pierce_2022}, as it was assumed that the relation would not change significantly with the inclusion of the $z<0.05$ sources. The stellar mass estimates for the 30 low redshift ($z<0.05$) targets are presented in Table~\ref{tab:sample_info_full}, whereas those for the higher redshift sample are available in \citetalias{pierce_2022}. However, three sources in the $0.05 < z < 0.3$ sample did not have XSC or PSC magnitudes available and therefore do not have stellar mass estimates. The stellar masses for the 10 3CR/2Jy objects were obtained from \cite{Bernhard_2022}, who followed the same methods, and were converted to our assumed cosmology. Overall, 109 of the 112 3CR sources have stellar mass estimates.

\subsection{Control matching procedure}
\label{control_matching}

The control galaxies were selected from the MPA-JHU value-added catalogue, containing derived properties for from 927 552 SDSS DR7 galaxy spectra, including spectroscopic redshifts and stellar mass estimates. A similar procedure to that detailed in \citetalias{pierce_2022} was followed to obtain a suitable sample of control galaxies. However, with the addition of the low-\textit{z} sample, some changes were made to the method.

\citetalias{pierce_2022} matched precisely in redshift to their 3CR sample, that covered a much larger redshift range ($0.05 < z < 0.3$). However, given that a much smaller volume of space was covered, and the stellar masses of the 3CR galaxies are typically high (median of $\log(M_\ast/M_\odot) = 11.3$ for the $z<0.05$ portion of sample), there were insufficient control galaxies for precise redshift matching to the $z<0.05$ 3CR objects. \citetalias{pierce_2022} found no significant trend in galaxy disturbance rate with redshift for their samples of active galaxies and matched controls. Therefore, we relaxed the redshift constraints when selecting control galaxies, allowing any galaxy with $z<0.3$ to be considered as a potential control. As we show in Section~\ref{mass_redshift} and Fig.~\ref{fig:stellar_mass_z_proportions}, the lack of a trend in galaxy disturbance rate with redshift holds for the 3CR and control samples selected for this study, so precise redshift matching is not required. However, \citetalias{pierce_2022} did find a relationship between the stellar mass and the disturbance rate in their sample of active and control galaxies. Therefore, the galaxies were matched precisely in stellar mass following their approach, which required the control galaxies to meet both of these conditions: 

\begin{enumerate}[leftmargin=*]
    \item(log($M_*$/$M_{\odot}$) + $\sigma$)$_{\rm control}$ $>$ (log($M_*$/$M_{\odot}$) $-$ $\sigma$)$_{\rm target}$\,;
    \item (log($M_*$/$M_{\odot}$) $-$ $\sigma$)$_{\rm control}$ $<$ (log($M_*$/$M_{\odot}$) $+$ $\sigma$)$_{\rm target}$\,.
\end{enumerate}

To reduce the otherwise large number of new images needed for the classification, we used all the control galaxies selected for the 3CR sample in \citetalias{pierce_2022}. However, with the addition of the new low-$z$ sample, the number of high mass galaxies increased relative to the controls. This caused a mismatch between the mass distributions of the 3CR and control galaxies. Hence, it was necessary to select more high mass control galaxies.

We utilised our more recent observations from the INT/WFC, taken as part of a long-standing observation programme, to study the triggering mechanisms of nearby AGN of various types, increasing the sky area over which control galaxies could be selected. The control catalogue filtering and mass matching process detailed above was followed and all the control galaxies from all the INT/WFC fields were identified. 

Where possible, the three closest control galaxies in stellar mass to each 3CR galaxy which met the conditions outlined above, were selected, not allowing repeat selections. This ensured that the majority of the 3CRs were matched to three unique control galaxies and the mass distributions of the two samples would be as close as possible. 

Of the 109 3CR galaxies which had stellar mass estimates available, 82 had three unique matches and one had a single control match (3C 438) from \citetalias{pierce_2022}, yielding 247 controls selected from this source. For the remaining 26 3CR galaxies, where possible, the three closest unique controls in stellar mass were selected from the pool of new controls found in the new INT/WFC fields. Of these 26 3CR galaxies, there were four which failed to be matched to any control galaxies (3C 83.1, 3C 129.1, 3C 130, 3C 410), one with two matches (3C 323.1), and one with a single match (3C 78). No further controls were found for 3C 438. Upon visual inspection, a further three controls could not be used due to image issues (e.g. bad image regions, defects), leaving the galaxies 3C 296, 3C 111 and 3C 465, with only two unique control matches. Therefore, a total of 60 new controls were selected. The lack of control matches for some of the 3CR targets was due to their high stellar masses, with stellar mass estimates in the range $11.7\leq\log(M_*/M_\odot)\leq12.7$.

\begin{figure}
    \centering
    \includegraphics[width=\linewidth]{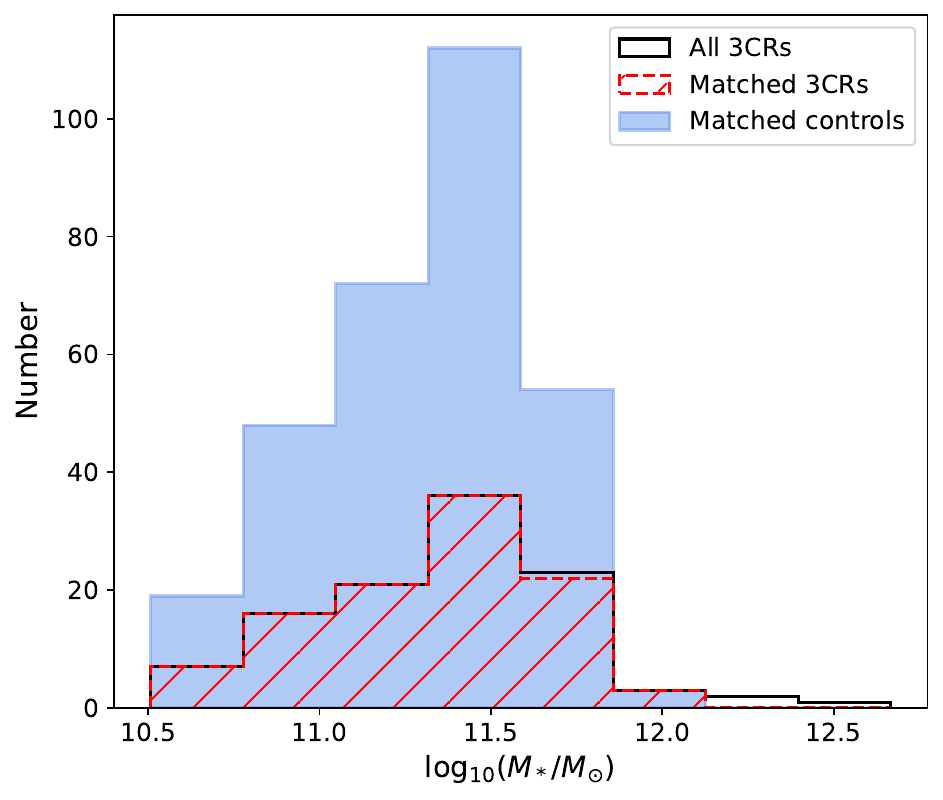}
    \caption{Stellar mass distribution of the 109 3CR sources which had stellar mass estimates, alongside the matched control sample distribution. The stellar mass distribution for the full 3CR sample, including those without control matches, is also shown.}
    \label{fig:stellar_mass}
\end{figure}

Overall, a total of 307 controls were selected, with 3 unique matches found for 91 per cent of the sample (99 of 109 targets), and at least 1 match for 96 per cent of the sample (105 of 109 targets). The stellar mass distributions of the 3CR sample and the matched control sample is shown in Fig.~\ref{fig:stellar_mass}. A two-sample Kolmogorov–Smirnov test provides insufficient evidence to reject the null hypothesis that the matched 3CRs and controls are drawn from the same underlying distribution (test statistic, $D=0.073$; $p$-value, $p=0.760$). The distributions are similar, apart from the high mass end (log($M_*$/$M_{\odot})$~>~11.7), where there is a paucity of control galaxies.

\subsection{Online classification interface - Zooniverse}
\label{zooniverse}

The Zooniverse Project Builder at \href{http://www.Zooniverse.org}{Zooniverse.org} \citep{Lintott_2008, Lintott_2011} was used to create an interface for the morphological classifications of all the galaxies included in the project, as in \citetalias{pierce_2022}. This method proved effective, both in obtaining the morphological classifications of the galaxies, but also ensuring that biases were not introduced during the process. 

Nine researchers (all authors except SM, RH, CRA and JR) were presented with images of the 3CR galaxies and the control galaxies randomly, and asked a series of multiple choice questions about their optical morphologies. To avoid introducing biases, the classification was done blindly, with no additional information given about the galaxy displayed (e.g. target name, AGN or control, stellar mass, redshift, optical luminosity). 

In the interface, the classifiers were presented with two images of each object with different contrast levels to show details of each target's morphology (see Fig.~\ref{dist_not_dist_examples}). This method allowed standardisation of what the classifiers were presented within the interface and, therefore, reduced the subjectivity of the classifications. Unlike prior studies where contrast manipulation was more flexible \citep[e.g.][]{ramos_almeida_2011, ramos_almeida_2012, Bessiere_2012, Ellison_2019, Pierce_2019}, our method provides a more controlled and uniform analysis of the sample. One image was high contrast, to aid with identification of any high surface brightness tidal features and overall galaxy morphology; and the other low contrast, to identify any low surface brightness features. The contrast levels were chosen manually for each image, to ensure the best representation of the target and the wider field. 

For the majority of objects, each image was a fixed size of 200 kpc $\times$ 200 kpc, determined at the redshift of the target\footnote{Three of the galaxies in our sample (3C~270, 3C~272.1 and 3C~274) are too nearby for redshift to be an accurate distance indicator. Instead, we use a median of redshift-independent distances measured after the year 2000, indicated on the NASA/IPAC Extragalactic Database (NED; \url{https://ned.ipac.caltech.edu/}). Distance measurement methods include the Faber-Jackson relation \citep{Faber_Jackson_1976}, Type 1a Supernovae light curves \citep[e.g.][]{Perlmutter_1999}, globular cluster luminosity functions \citep[e.g.][]{Hanes_1977} and surface brightness fluctuations \citep[e.g.][]{Tonry_Schn_1988, Tonry_1990}. Median distances of 32.4 Mpc, 31.2 Mpc and 31.1 Mpc were calculated for 3C~270, 3C~272.1 and 3C~274, respectively.}, and was centred on the galaxy. However, due to their low redshift, two of the objects (3C~272.1 and 3C~274) required a smaller image size of 100 kpc $\times$ 100 kpc. Two scale bars of 10 kpc were also displayed on each image to help with multiple nucleus classification (see below). The images for all the low-\textit{z} 3CRs are presented in Appendix~\ref{stamp_images}, while those for the higher redshift portion of the sample are presented in the supplementary material of \citetalias{pierce_2022}. The images of the 10 3CR/2Jy objects are presented in \cite{ramos_almeida_2011}.

For consistency with our previous work, the questions asked to the classifiers in the interface were the same as in \citetalias{pierce_2022}, as summarised here. The first question asked `Does this galaxy show at least one clear interaction signature?', with the options: (i) `Yes'; (ii) `No'; or (iii) `Not classifiable (e.g. image defect, bad region/spike from saturated star)’. A majority threshold was used to determine the outcome of this question. A galaxy was considered disturbed if five or more votes were for `Yes', and not disturbed if five or more votes were for `No'. All other vote distributions, including any amount of votes for `Not classifiable', were regarded as uncertain. To allow results for cases which some classifiers had voted as uncertain, if more than half of the researchers had classified the object (i.e. 5 votes minimum in total for the options `Yes' and `No'), the majority vote was taken as the classification. If a majority was still not achieved, objects were given an `uncertain' classification.

If the first question was answered with `Yes', the classifiers were then asked `What types of interaction signature are visible?’. Multiple selections from the following list of interaction signatures were allowed: (i) Tail (T), (ii) Fan (F), (iii) Shell (S), (iv) Bridge (B), (v) Amorphous halo (A), (vi) Irregular (I), (vii) Multiple nuclei (2N, 3N...) within 10 kpc, (viii) Dust lane (D), (ix) Tidally interacting companion (TIC). This classification scheme is the same as that in \citetalias{pierce_2022} (see their paper for more detailed explanation of the individual interaction signatures), and also largely consistent with \cite{ramos_almeida_2011, ramos_almeida_2012} and \cite{Bessiere_2012}\footnote{Note that the 'Tidally interacting companion' classification was not available in \cite{ramos_almeida_2011,ramos_almeida_2012} or \cite{Bessiere_2012}.}.

The final question asked the classifiers `On first impression, what is the morphological type of the galaxy?', with the options: (i) `Spiral/disk'; (ii) `Elliptical; (iii) `Lenticular'; (iv) `Merger (too disturbed to classify)'; or (v) `Unclassifiable (due to image defects, \textit{not} merger'. If the classifiers answered the first question with `No', they were taken to answer this question straight away. 

To aid with the classifications, a tutorial was provided, which the classifiers could access in the Zooniverse project. Example images of all the interaction signatures and the galaxy morphological type were provided, alongside more detailed descriptions. 

We emphasize that all 72 3CR sources from the original \citetalias{pierce_2022} study were reclassified in this work and good agreement was found with the original study. Despite the subjectivity of the classification and the fact that five of the classifiers were different from the original study, only 3 of the 72 sources ($\sim$4 per cent) differed in their classifications. We also reclassified 247 of the galaxies from their control sample. Of these 247 control galaxies, 29 differed in their classifications ($\sim$12 per cent). However, of these 29, 18 ($\sim$62 per cent) were classified as uncertain in the original study. Due to the odd number of classifiers in our study (9 classifiers), it follows that a majority can be more easily achieved than for an even number of classifiers \citepalias[8 classifiers in][]{pierce_2022}, therefore, reducing the number of uncertain classifications. Overall, the small differences found in the classifications demonstrates the robustness of this classification method.

\section{Analysis and results}
\label{analysis+results}

The main focus of our work was to determine whether there were significant differences in the triggering mechanisms of the different sub-types of radio AGN, and investigate the importance of galaxy mergers and interactions in this context. Morphological classifications were obtained from the nine classifiers via the online interface for the 419 3CR and control galaxies included in the project. Full detailed classification results for the 112 3CRs included in this study are presented in Appendix~\ref{classification_results}. 

Fig.~\ref{dist_not_dist_examples} shows example images of some of the low-\textit{z} 3CR galaxies which were classified as disturbed and not disturbed from the online interface classification.  

As well as for the first question asked to the classifiers (see Section~\ref{zooniverse}), we also use a majority threshold to determine the outcome of the third question regarding the host galaxy morphology. We find that 93$^{+2}_{-4}$ per cent of our 3CR sample are hosted in elliptical galaxies. This is consistent with previous work that found the most powerful radio AGN to be hosted by massive early-type galaxies \citep[e.g.][]{Matthews_1964, pierce_2022}. 

\begin{figure*}{}
    \centering
    \includegraphics[width=0.9\linewidth]{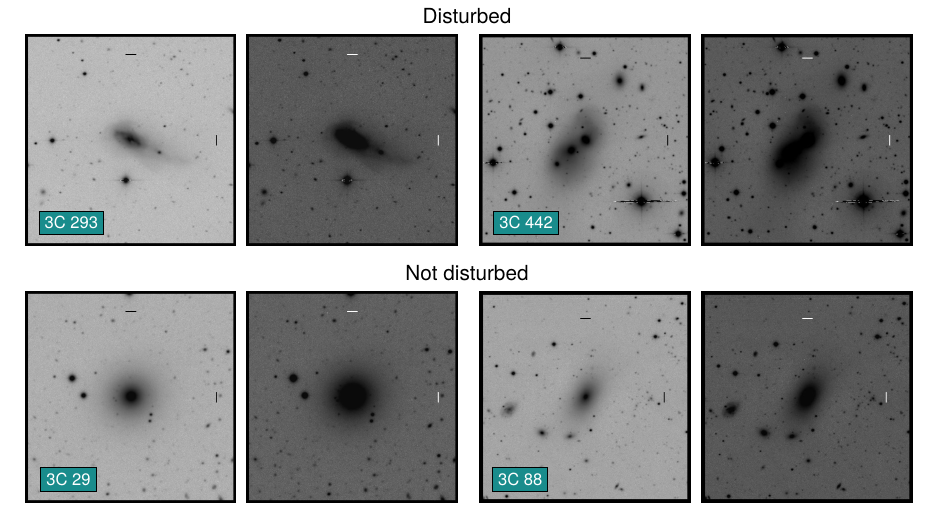}
\caption{Examples of the 3CR galaxies classified as disturbed (top panel) and not disturbed (bottom panel) from the online interface classification (images for all the low-$z$ ($z<0.05$) are presented in Appendix~\ref{stamp_images}). Each image is 200 kpc $\times$ 200 kpc, centred on the target galaxy, with 10 kpc scale bars to aid with classification (see Section~\ref{zooniverse}). The high contrast (left of pair) and low contrast (right of pair) images are presented.}
\label{dist_not_dist_examples}
\end{figure*}


\subsection{Proportions}
\label{proportions}

\begin{figure}
    \centering
    \includegraphics[width=\linewidth]{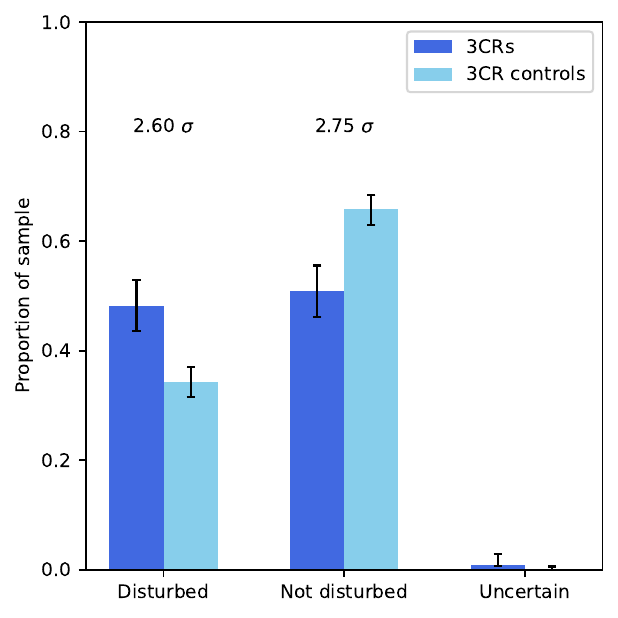}
    \caption{Proportions of the whole 3CR sample which were classified as disturbed, not disturbed, or uncertain, alongside the matched control sample.}
    \label{fig:proportions_all_3CRs}
\end{figure}

The proportions of the full 3CR sample and controls classified through the online interface as disturbed, not disturbed, or uncertain are presented in Fig.~\ref{fig:proportions_all_3CRs} and Table~\ref{tab:table_props_all_3CR}. In each case, the AGN sample is presented alongside its stellar mass matched control sample, with the significance of the differences between the two samples estimated using bootstrap resampling, involving a minimum of 10,000 resamples. In those cases where one of the sample sizes was zero, the significance of the difference is not presented, as these statistics are inherently inaccurate. We followed the Bayesian approach of \cite{Cameron_2011}, using the quantiles of the beta distribution to determine the proportion uncertainties\footnote{The proportion uncertainties estimated using the \cite{Cameron_2011} approach were consistent with those derived from bootstrapping, when the proportions were not zero.}. 

\begin{table*}
\def\arraystretch{1.5}
    \centering
    \caption{The proportions of the whole 3CR sample ($z<0.3$) and the sub-types in the sample which were classified as disturbed, not disturbed, or uncertain, presented alongside their matched control samples. All the proportions are presented as percentages. The number of objects ($N$) in each sample is also given.}
    \label{tab:table_props_all_3CR}
    \begin{tabular}{lccccccc}
    \hline
        & &\multicolumn{2}{c}{Disturbed (\%)} & \multicolumn{2}{c}{\makecell{Not dist. (\%)}} & \multicolumn{2}{c}{Uncertain (\%)} \\ 
        & \textit{N} & AGN & Cont. & AGN & Cont. & AGN & Cont. \\ 
    \hline
    \makecell[l]{Full 3CR \\} & 112 & 48 $\pm$ 5 & 34 $\pm$ 3 & 51 $\pm$ 5 & 66 $\pm$ 3 & 1 $^{+2}$ & 0 $^{+1}$ \\
    
    \makecell[l]{HERGs\\ } & 53 & 62 $^{+6}_{-7}$ & 31 $\pm$ 4 & 36 $^{+7}_{-6}$ & 69 $\pm$ 4 & 2 $^{+4}_{-1}$ & 0 $^{+1}$ \\
    
    \makecell[l]{LERGs\\ } & 58 & 36 $^{+7}_{-6}$ & 37 $\pm$ 4 & 64 $^{+6}_{-7}$ & 63 $\pm$ 4 & 0 $^{+3}$ & 0 $^{+1}$ \\ 

    \makecell[l]{FRIs\\ } & 23 & 43 $^{+10}_{-9}$ & 32 $^{+7}_{-6}$ & 57 $^{+9}_{-10}$ & 68 $^{+6}_{-7}$  & 0 $^{+7}$ & 0 $^{+3}$ \\ 
    
    \makecell[l]{FRIIs \\} & 87 & 49 $\pm$ 5 & 33 $\pm$ 3 & 49 $\pm$ 5 &  67 $\pm$ 3 & 1 $^{+3}$ & 0 $^{+1}$ \\ 

    \makecell[l]{FRI LERGs \\} & 23 & 43 $^{+10}_{-9}$ & 32 $^{+7}_{-6}$ & 57 $^{+9}_{-10}$ & 68 $^{+6}_{-7}$  & 0 $^{+7}$ & 0 $^{+3}$ \\   

    \makecell[l]{FRII HERGs \\} & 53 & 62 $^{+6}_{-7}$ & 31 $\pm$ 4 & 36 $^{+7}_{-6}$ & 69 $\pm$ 4 & 2 $^{+4}_{-1}$ & 0 $^{+1}$ \\  
    
    \makecell[l]{FRII LERGs \\} & 33 & 30 $^{+9}_{-7}$ & 36 $\pm$ 5 & 70 $^{+7}_{-9}$ & 64 $\pm$ 5 & 0 $^{+5}$ & 0 $^{+2}$ \\ 

    \makecell[l]{3CR quasars \\} & 26 & 65 $^{+8}_{-10}$ & 28 $^{+6}_{-5}$ & 35 $^{+10}_{-8}$ & 72 $^{+5}_{-6}$ & 0 $^{+7}$ & 0 $^{+3}$ \\ 
    
    \hline
    \end{tabular}
\end{table*}


Considering the sample as a whole, the proportions of the disturbed (48$\pm$5 per cent) and non-disturbed 3CR objects (51$\pm$5 per cent) are similar. However, the 3CRs do show evidence for an excess in disturbed morphologies when compared to the matched control sample, with the differences significant at the $\sim$2.6$\sigma$ level. When comparing to the results for the higher redshift part of the 3CR sample and matched controls from \citetalias{pierce_2022}, it can be seen that the significance of the differences between the control and radio galaxy populations in this work is lower, with a lower proportion of the 3CR sample classified as disturbed. However, this is likely due to the inclusion of the low-\textit{z} sample, consisting largely of LERGs, which display a lower proportion of disturbed morphologies (see Section~\ref{HERG_LERG}).

The proportions of the disturbed, not disturbed, and uncertain classifications for different radio AGN sub-types in the sample are discussed in the following sections, where we determine whether there are statistically significant differences between them.

\subsubsection{HERGs and LERGs}
\label{HERG_LERG}

\begin{figure*}
    \centering
    \includegraphics[width=\linewidth]{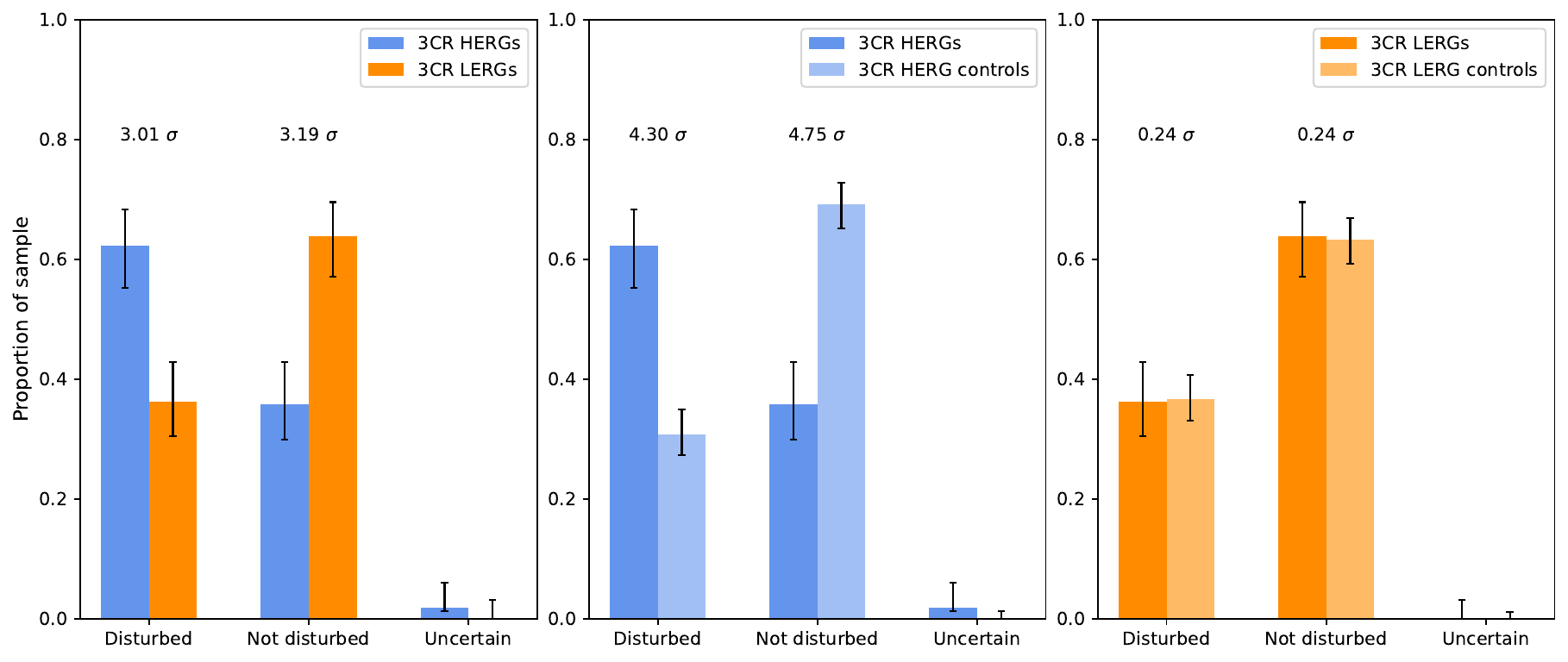}
    \caption{Proportions of the 3CR HERGs and LERGs classified as disturbed, not disturbed, or uncertain. The HERGs and LERGs are presented alongside each other (first panel), and with their respective matched control samples (second and third panel).} 
    \label{fig:proportions_HERG/LERG}
\end{figure*}

When considering the HERG and LERG sub-types in the 3CR sample, the proportions for the disturbed, not disturbed and uncertain objects, alongside those for their matched control samples, are presented in Fig.~\ref{fig:proportions_HERG/LERG} and in Table~\ref{tab:table_props_all_3CR}.

Our results show significant differences ($>$3$\sigma$) between the sub-types of the powerful radio AGN population. The 3CR HERGs show a strong preference for disturbed morphologies, with $62^{+6}_{-7}$ per cent showing clear signs of disturbance, in contrast to only $36^{+7}_{-6}$ per cent of the 3CR LERGs. The difference between the two populations is significant at the $>$3$\sigma$ level. The 3CR HERGs also exhibit $>$4$\sigma$ differences with their matched controls, whereas the 3CR LERGs were consistent with their matched controls.

We find good agreement between our results and those from \citetalias{pierce_2022}, who found $66^{+7}_{-8}$ per cent of the 3CR HERGs in their sample were hosted in morphologically disturbed galaxies, compared with only $37^{+9}_{-8}$ per cent of the LERGs. The difference between the HERG and LERG proportions in their study was, however, only significant at the 2.4$\sigma$ level. This was largely due to the small numbers of LERGs in their sample. With the inclusion of the low-$z$ 3CR sample in this work, we improve upon this result, finding the difference between the HERG and LERG proportions is significant at the $>$3$\sigma$ level. 

\subsubsection{FRIs and FRIIs}
\label{fr1_fr2}

It was not possible in our previous study to separately investigate how the disturbance rates depend on radio morphology (i.e. whether FRI or FRII), as only a few sources in the \citetalias{pierce_2022} 3CR sample are FRIs. With the inclusion of the low-\textit{z} sample, we increased the number of 3CR FRI sources from 3 to 23 objects, allowing us to investigate the incidence of tidal features in relation to radio morphology. Although the majority of the FRIs are concentrated at the lower redshift end of our sample ($z<0.05)$ compared with the FRIIs (see Appendix~\ref{fr1_fr2_dists}), given the lack of dependence of disturbance rate with redshift (see Section~\ref{mass_redshift}) we do not expect this to have any major impacts on our results.

\begin{figure*}
    \includegraphics[width=\linewidth]{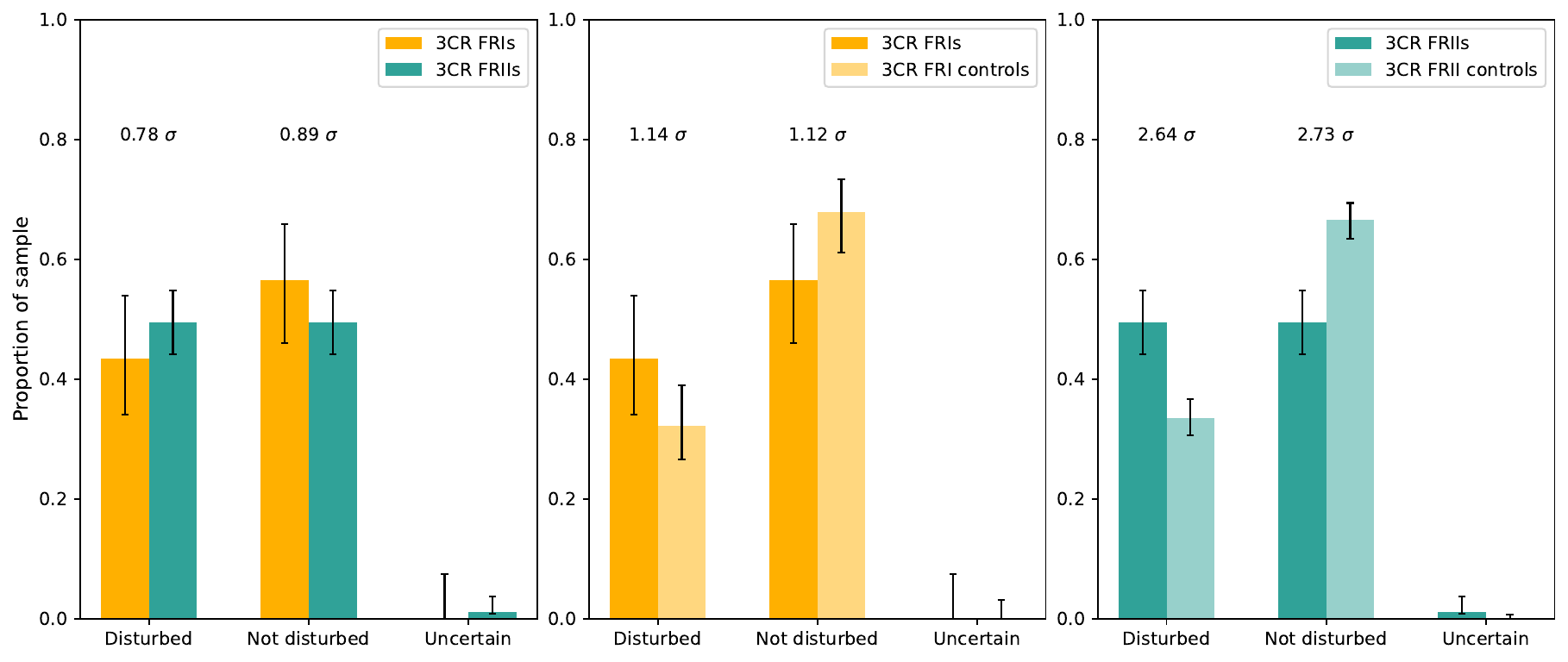}
    \caption{Proportions of the 3CR FRIs and FRIIs classified as disturbed, not disturbed, or uncertain. The two samples are presented alongside each other (first panel), and with their respective matched control samples (second and third panel).}
    \label{fig:proportions_FRI/FRII}
\end{figure*}

The proportions of disturbed, not disturbed and uncertain classifications for the 3CR FRI and FRII sub-types are presented in Fig.~\ref{fig:proportions_FRI/FRII}, alongside their matched controls. The measured proportions for each sample are presented in Table~\ref{tab:table_props_all_3CR}. 

The 3CR FRIIs display a marginally more disturbed morphology than the 3CR FRIs, however the differences in the proportions of disturbed and not disturbed classifications for the two populations are not statistically significant. This suggests that, unlike the optical classification, is not a strong, separate dependence of the triggering mechanisms on the radio morphology. However, when compared to their respective matched control samples, the 3CR FRIIs are slightly more likely to be hosted in galaxies with disturbed morphologies, exhibiting differences significant at the $\sim$2.6$\sigma$ level, compared to only $\sim$1.1$\sigma$ for the FRIs and their controls. This is likely driven by the fact that FRIIs are more commonly associated with HERGs (61 per cent of the FRIIs in our sample are HERGs), which are predominantly more disturbed than the LERGs with respect to their matched control samples (see Section~\ref{HERG_LERG}). 

\begin{figure}
    \centering
    \includegraphics[width=\linewidth]{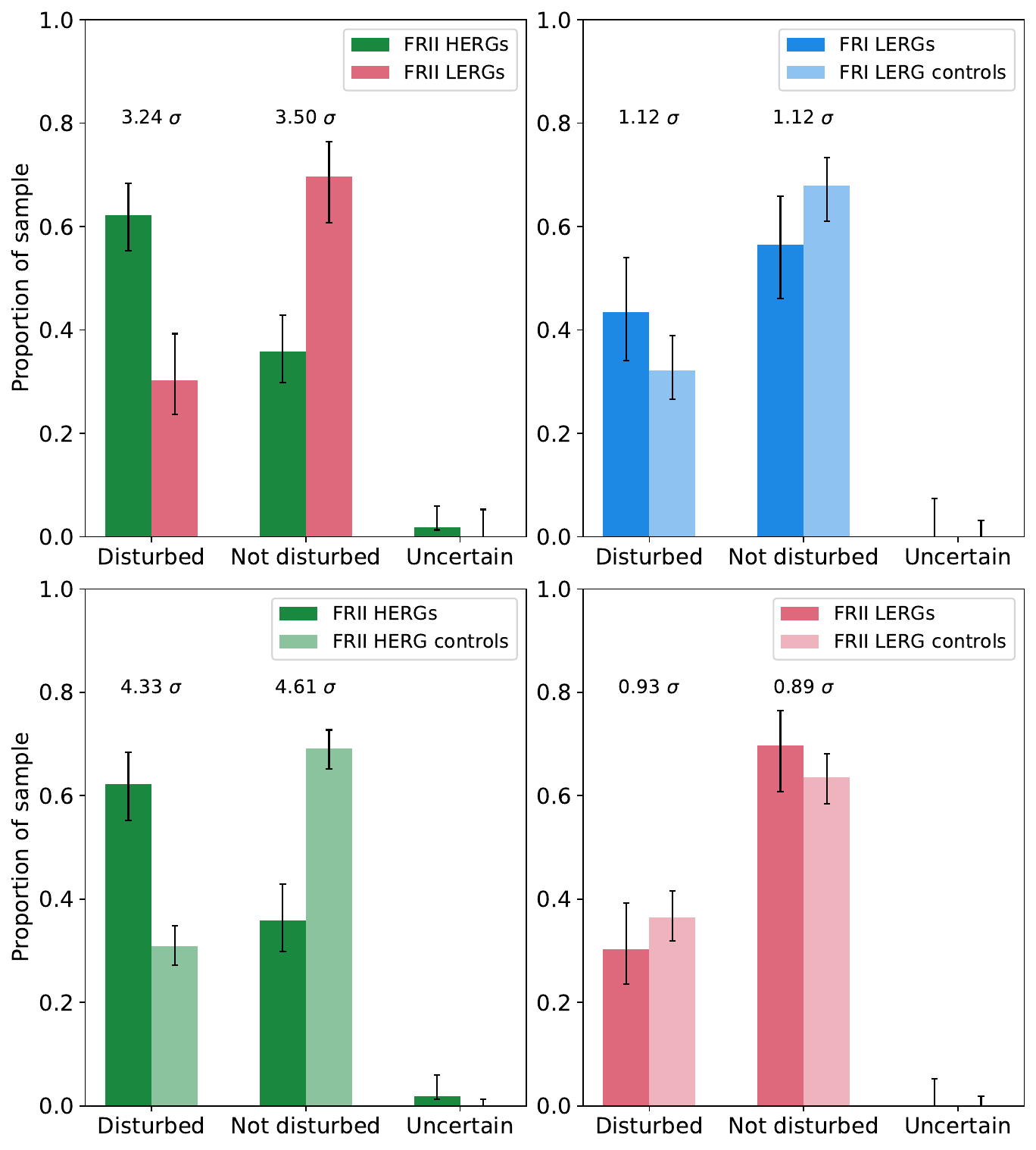}
    \caption{The proportions of disturbed, not disturbed and uncertain classifications for FRII HERGs, FRII LERGs and FRI LERGs. The FRII HERGs and LERGs are presented alongside each other, and with their matched control samples. The FRI LERGs are presented alongside their matched controls.}
    \label{fig:FR_props_HERG/LERG+conts}
\end{figure}

To further explore any triggering relationship with radio AGN sub-type, we also consider the disturbance rates when combining the radio and optical classifications of our 3CR sample. We discuss the broader implications of our results for the combined classifications in Section~\ref{fr1_fr2_discuss}.

Fig.~\ref{fig:FR_props_HERG/LERG+conts} presents the proportions of disturbed, not disturbed and uncertain classifications for the FRII HERGs compared with the FRII LERGs (all the FRI galaxies in our sample are LERGs, so their proportions are not presented here). The proportions and associated uncertainties are also shown in Table~\ref{tab:table_props_all_3CR}. The FRII HERGs show an excess in disturbed morphologies, significant at the $3.24\sigma$ level, when compared to the FRII LERGs; indeed, the FRII LERGs show the lowest disturbed proportion of all the radio AGN sub-types (30$^{+9}_{-7}$ per cent). This is consistent with the HERGs contributing to the increased disturbance rate observed in our FRII sample, as discussed above.

The proportions of disturbed, not disturbed and uncertain classifications for the FRI LERGs, FRII LERGs and FRII HERGs are also presented alongside their respective matched controls in Fig.~\ref{fig:FR_props_HERG/LERG+conts}. This also supports the picture that the FRII HERGs exhibit excesses in their disturbed morphologies, showing a $\sim$4$\sigma$ difference with respect to their matched controls, driving the excess seen in the FRII sample as a whole. In contrast, both the FRII LERGs and FRI LERGs show only $\sim$1$\sigma$ differences.

\subsubsection{3CR quasars}
\label{3CR_quasars}

\begin{figure}
    \centering
    \includegraphics[width=\linewidth]{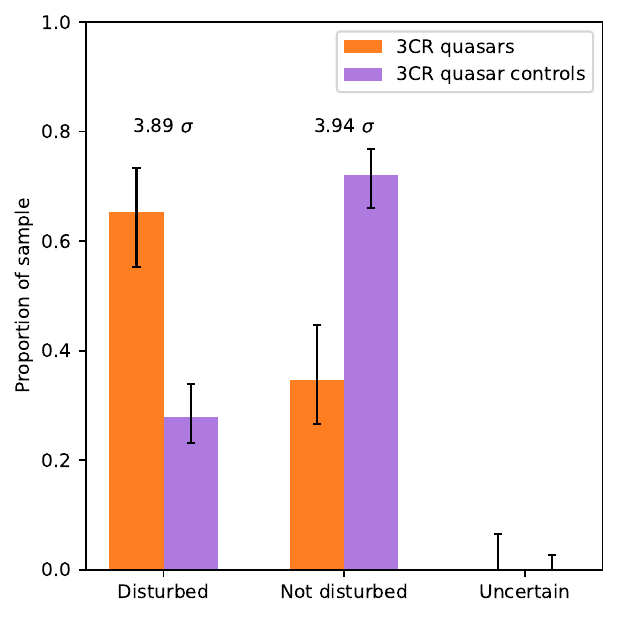}
    \caption{The proportions of disturbed, not disturbed and uncertain classifications for the quasar-like (L$_{\mathrm{[OIII]}}\geq10^{35}$~W) portion of the 3CR sample and their matched control sample.}
    \label{fig:3CR_quasar_proportions}
\end{figure}

Given the evidence that the rate of morphological disturbance increases with AGN luminosity \citepalias{pierce_2022}, it is interesting to consider the morphological properties of the most luminous, quasar-like, objects in the 3CR sample. Fig.~\ref{fig:3CR_quasar_proportions} shows the proportions of disturbed, not disturbed and uncertain classifications for the quasar-like AGN in our 3CR sample (L$_{\mathrm{[OIII]}}\geq10^{35}$~W), alongside their matched controls. 

The 3CR quasars show a clear preference for disturbed morphologies, with $65^{+8}_{-10}$ per cent classified as disturbed, compared with $28^{+6}_{-5}$ per cent for their matched control sample -- an excess of $\sim$4$\sigma$. This supports the proposal that galaxy mergers and interactions provide the dominant triggering mechanism for the highest luminosity AGN. Our results are also in line with the findings from \citetalias{pierce_2022} who showed that there was an increasing importance of galaxy mergers and interactions with higher [OIII] emission-line luminosities. They also found $67^{+8}_{-11}$ per cent of their 3CR sample with quasar-like luminosities were hosted in disturbed galaxies, in very good agreement with our results. All of these objects were reclassified in this work, once again demonstrating the robustness of the classification method. 

We also find consistent results with those for the radio-quiet Type 2 quasar sample from \cite{Pierce_2023}, who found that $65^{+6}_{-7}$ per cent of the Type 2 quasars were hosted in galaxies with clear signs of morphological disturbance. In addition, the moderate redshift sample ($0.3 < z < 0.41$) sample of Type 2 quasars from \cite{Bessiere_2012} also showed similar rates of disturbance ($75\pm20$ per cent). 

\subsection{Interaction signatures and merger stage}
\label{pre_post_coal}

To characterise the type and stage of the galaxy mergers and interactions, the classifiers were asked `What types of interaction signature are visible?’ (see Section~\ref{zooniverse}). A majority threshold of 5 of 9 votes for `Yes' to the first question in the interface had to be met, to ensure that only disturbed galaxies were included. Then, this majority had to have voted for that corresponding interaction signature for it to be considered a secure classification.

The aim was to determine whether the galaxy was in the early or late stages of a merger, i.e. before (`pre-coalescence') or after (`coalescence/post-coalescence') the coalescence of the galaxy nuclei. In line with \citetalias{pierce_2022}, and also with \cite{ramos_almeida_2011} and \cite{Bessiere_2012}, the interaction signatures `Multiple nuclei', `Bridge' and `Tidally interacting companion' were taken as representative of an early-stage/pre-coalescence merger or interaction. The rest of the classifications were considered to be indicative of a late-stage or coalescence/post-coalescence (late-stage/post-coalescence hereafter, for conciseness) merger. Note that, because we do not have in-depth information regarding the stellar and gas dynamics in the systems, we cannot say for certain that the objects classified as pre-coalescence will eventually coalesce in the future. 

\begin{figure}
    \centering
    \includegraphics[width=\linewidth]{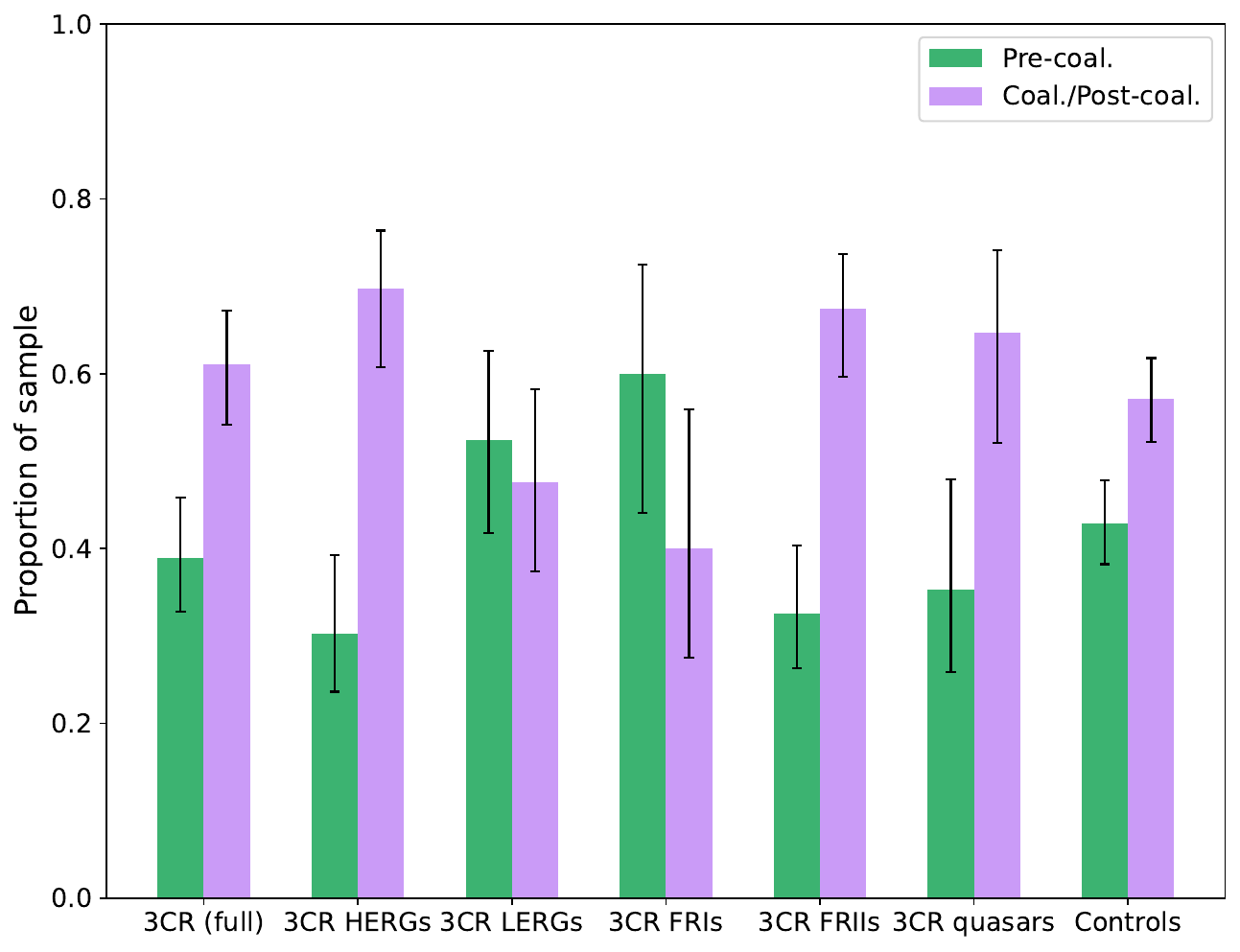}
    \caption{The proportion of disturbed galaxies classified as pre-coalescence or post-coalescence for the total 3CR sample, as well as the different sub-types. The proportions for the full control sample are also presented. Only galaxies with secure classifications of the pre- or post-coalescence signatures were considered (refer to the main text). The proportions were determined relative to the total number of these classifications for each sample, and are presented in Table~\ref{tab:pre_post_coal_table}.}
    \label{fig:pre_post_coal_fig}
\end{figure}

Fig.~\ref{fig:pre_post_coal_fig} shows the proportions of the disturbed 3CR samples and control galaxies which had secure classifications indicating pre- or post-coalescence interactions. The proportions for each sample are also presented in Table~\ref{tab:pre_post_coal_table}. We do not find any significant differences in the proportions of pre-/post-merger objects between the different radio AGN sub-types, because of our small number statistics (only the disturbed objects are being considered here), and resulting large uncertainties on the proportions. 

When considering the whole 3CR sample, late-stage mergers and interactions are preferred, with 61$^{+6}_{-7}$ per cent of disturbed objects favouring post-coalescence interactions compared with only 39$^{+7}_{-6}$ per cent for pre-coalescence interactions. Our results are in good agreement with, and also reduce the proportion uncertainties of \citetalias{pierce_2022} where 65$^{+9}_{-13}$ per cent and 35$^{+15}_{-9}$ per cent of their 3CR sample were classified as representative of late- and early-stage interactions, respectively. Moreover, our findings are consistent with the results of \cite{ramos_almeida_2011} for the 2Jy sample, where 65 $\pm$ 11 of disturbed galaxies were classified as post-coalescence and 35 $\pm$ 11 per cent as pre-coalescence. 

The preference for late-stage interactions also seems to hold consistent for most of the 3CR sub-types, except for the LERGs and FRIs, where the proportions are consistent with being equal -- unsurprisingly as all the FRIs in our sample are LERGs. In contrast, \cite{Pierce_2023} found the majority of their disturbed radio-quiet Type 2 quasar sample were hosted in pre-coalescence systems (61$^{+8}_{-9}$ per cent). 

\begin{table}
\centering
\small
\renewcommand{\arraystretch}{1.2}
\caption{The proportion of disturbed galaxies which displayed secure classifications indicting pre-coalescence or post-coalescence mergers and interactions, as seen in Figure~\ref{pre_post_coal}. All proportions are presented as percentages, and given relative to the combined number of secure pre- and post-coalescence classifications for each sample. The number of objects ($N$) in each sample is also given.}
\label{tab:pre_post_coal_table}
\begin{tabular}{lccc}
\hline
& \textit{N} & Pre-coalescence & Post-coalescence \\ 
\hline
3CR (full)  & 54 & 39 $^{+7}_{-6}$  & 61 $^{+6}_{-7}$    \\
3CR HERGs   & 33 & 30 $^{+9}_{-7}$  & 70 $^{+7}_{-9}$  \\
3CR LERGs    & 21 & 52 $^{+10}_{-11}$  & 48 $^{+11}_{-10}$    \\
3CR FRIs & 10 & 60 $^{+13}_{-16}$  & 40 $^{+16}_{-13}$   \\
3CR FRIIs   & 43 & 33 $^{+8}_{-6}$  & 67 $^{+6}_{-8}$  \\
3CR quasars  & 17 & 35 $^{+13}_{-9}$  & 65 $^{+9}_{-13}$   \\
All controls  & 105 & 43 $\pm$5  & 69 $\pm$5  \\
\hline
\end{tabular}
\end{table}

\subsection{Relationship with stellar mass and redshift}
\label{mass_redshift}

For any given seeing conditions, the effective spatial resolution of the observations in kpc will be worse at higher redshifts; also, the higher redshift objects will suffer increased levels of surface brightness dimming: 1.1 mag more at redshift $z=0.3$ compared with zero redshift. Therefore, we may expect to see a trend in the disturbance rate with the redshift. In addition, \citetalias{pierce_2022} showed an increase in the disturbance rate towards higher stellar masses. Such a trend is also seen in morphological studies of non-active galaxies \citep[e.g.][]{Desmons_2025}. With the addition of the low-\textit{z} sample and the 10 3CR/2Jy objects, we investigate the relationship between the disturbance rate, stellar mass, and redshift across our whole 3CR and control sample. 

\begin{figure}
    \centering
    \includegraphics[width=\linewidth]{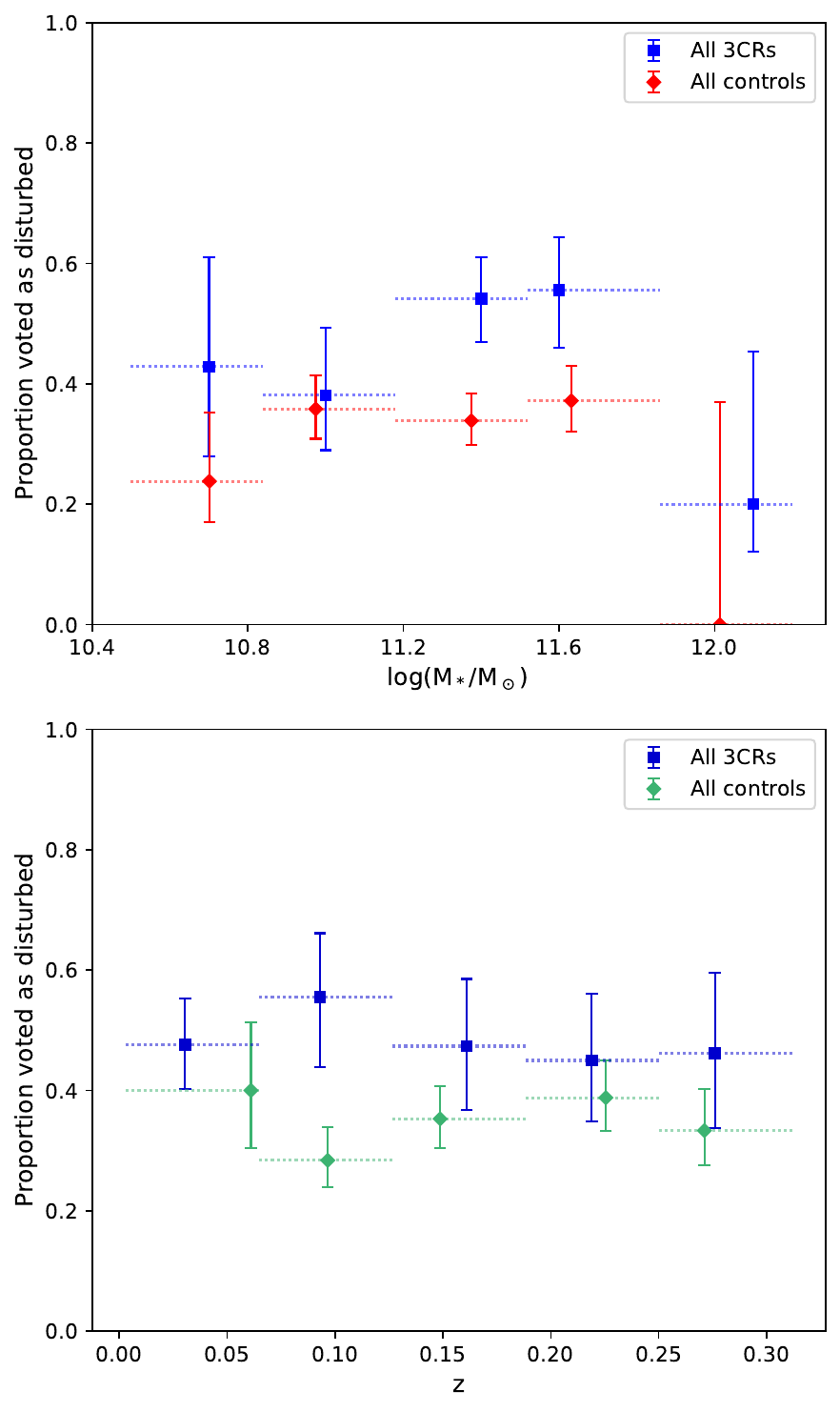}
    \caption{The proportions of the 3CR sample and the control sample which were voted as disturbed with stellar mass and redshift. The proportions of the 3CR and control sample are plotted at the median stellar masses and redshifts for each bin.}
    \label{fig:stellar_mass_z_proportions}
\end{figure}

Fig.~\ref{fig:stellar_mass_z_proportions} shows the proportion of 3CR galaxies voted as disturbed with stellar mass and redshift. Due to its high stellar mass in relation to the other galaxies in the sample, 3C 130 (log$(M_*/M_\odot)=12.7$) was not included in the plots.

Across most of the stellar mass and redshift range, the 3CR galaxies display a higher proportion of disturbed morphologies, as seen in Fig.~\ref{fig:stellar_mass_z_proportions}. This is consistent with the picture that the radio AGN are generally more disturbed than their matched controls, as seen in Fig.~\ref{fig:proportions_all_3CRs} and Table~\ref{tab:table_props_all_3CR}. However, we find no significant trend between the proportion of disturbed galaxies and the stellar mass, both for the 3CR and control sample. This excludes consideration of the highest mass bin due to the extremely low relative numbers, and resulting large proportion uncertainties, of both 3CR and control galaxies in this bin. Plotting the 3CR proportions with and without the inclusion of the unmatched 3CR objects in stellar mass brings no discernable difference in the proportions. Therefore, we do not expect any strong effects on our results for the 3CR and control sample disturbance rates, discussed in the previous sections, with the inclusion of these unmatched objects. 

At first sight, these results may seem inconsistent with previous studies that find evidence for a trend in disturbance rate with stellar mass \citepalias[e.g.][]{pierce_2022, Desmons_2025}. However, over the relatively small mass range covered by our sample, the increase in disturbance rate with stellar mass is mild in the previous studies. Therefore, given the relatively large error bars on our measured disturbance fractions, our results are consistent with the trend noted in the other studies, even if they do not by themselves provide strong independent evidence to support it.

It is also notable that we find no significant trend of the proportion of galaxies classified as disturbed with redshift, both for the 3CR and the control sample, confirming the previous result from \citetalias{pierce_2022}. Therefore, there is no strong evidence to suggest that the removal of the precise redshift constraint during the control matching procedure (see Section~\ref{control_matching}) will have any significant effect on our results. 

\section{Discussion}
\label{discussion}

In the following sections, we discuss our results in the context of how different radio AGN sub-types may be triggered. 

\subsection{HERG/LERG}
\label{herg_lerg_discussion}

From our analysis based on the classifications from the online interface, 62$^{+6}_{-7}$ per cent of the 3CR HERGs are classified as disturbed -- an excess of $\sim$4$\sigma$ in comparison to their matched controls. This is consistent with the disturbance rates found by \citetalias{pierce_2022}, where $66^{+7}_{-8}$ per cent of their HERGs exhibited clear signs of morphological disturbance, a difference significant at the 4.7$\sigma$ level relative to their matched control sample.
 
This trend of excess in disturbed morphologies for the radiatively-efficient AGN is also consistent with the results from \cite{ramos_almeida_2011}, in which $94^{+2}_{-7}$ of the 2Jy SLRGs showed signs of disturbance. However, although we find the same trend, our proportion is significantly lower. The classifications of the 2Jy sample were obtained based on images taken under better seeing conditions and using a more detailed, non-blind analysis, with full manipulation of the images allowed. In contrast, our method obtained classifications based on 2 images of different contrast levels. Of the 8 2Jy HERG/SLRG sources which were reclassified in our work, 5 were classed as disturbed (62.5 per cent), compared with 7 (87.5 per cent) in the original \cite{ramos_almeida_2011} paper. Although the numbers are small, this is consistent with the idea that our classification method using the online interface is significantly more conservative than that of \cite{ramos_almeida_2011}. However, we emphasize that our method provided a significant advantage by standardising the classifications across all the classifiers, thereby reducing the subjectivity. The reduced sensitivity to low surface brightness features using the online interface method, combined with the other factors discussed above, likely accounts for the lower rate of merger classifications in this work compared with \cite{ramos_almeida_2011}. 

We emphasize that, despite the significant overlap, the SLRG/WLRG and HERG/LERG classifications are not exactly the same \citep{Tadhunter_2016}. In light of this, we reviewed the proportions of disturbed, not disturbed and uncertain classifications of the 3CR sources using the alternative optical SLRG/WLRG classification scheme used in \cite{ramos_almeida_2011}. Our results are presented in Figure~\ref{fig:props_SLRG_WLRG} in Appendix~\ref{SLRG_WLRG}. We find the same trend for the SLRG/WLRG sub-types as for the HERG/LERG classification, with the radiatively-efficient SLRGs favouring disturbed morphologies (58$\pm 6$ per cent) compared to the radiatively-inefficient WLRGs (33$^{+8}_{-6}$ per cent). However, the results are less significant than for the HERG/LERG classification, with the difference between the measured proportions for the SLRG/WLRG significant at a lower level ($\sim$2.8$\sigma$). This suggests that the HERG/LERG classification may be a more effective method for identifying radiatively-efficient/inefficient AGN than the SLRG/WLRG scheme, assuming that radiatively-efficient objects are predominantly merger-triggered.

Considering the 3CR LERGs, 36$^{+7}_{-6}$ per cent are classified as disturbed from our analysis using the online interface classification method. This is in agreement with the disturbance rates in \citetalias{pierce_2022} and \cite{ramos_almeida_2011}, 37$^{+9}_{-8}$ per cent and 27$^{+16}_{-9}$ per cent, respectively. As we have a much larger sample of the radiatively-inefficient objects compared to these 2 studies, we obtain smaller proportion uncertainties. We also find no significant excess in relation to their matched control sample, consistent with \citetalias{pierce_2022} and \cite{ramos_almeida_2012}, and similar results to studies of predominantly intermediate radio luminosity LERGs. \cite{Gordon_2019} found evidence of disturbance in $28.7\pm1.1$ per cent of their sample of 282 low redshift ($z<0.07$), lower radio luminosity ($10^{21.7}<\mathrm{log(L_{1.4~GHz})}<10^{25.8}~\mathrm{W~Hz^{-1}}$) LERGs. Furthermore, \cite{Ellison_2015} determined their sample of intermediate radio luminosity LERGs, when compared to a sample of non-active control galaxies, displayed no significant enhancement in post-merger signatures or close pairs relative, when controlling for both the host galaxy and environmental properties. 

Overall, our results show significant differences ($>$3$\sigma$) between the optical morphologies of the HERG and LERG sub-types of the powerful radio AGN population. This provides clear evidence that galaxy mergers and interactions are an important triggering mechanism for the radiatively-efficient AGN, but much less so for the radiatively-inefficient AGN. It also ties in with other evidence for differences between the LERG and HERG host galaxies and environments. For example, a study by \cite{Bernhard_2022} of the powerful radio AGN in the 2Jy sample, covering redshifts $0.05<z<0.7$, found their WLRGs exhibited both lower star formation rates and cool ISM masses than the SLRGs. Since radiatively-efficient accretion processes are typically fuelled by reservoirs of cold gas \citep[e.g.][]{Hardcastle_2007}, the lower cool ISM content in the WLRGs suggests a reduced rate of such accretion in these objects. This is consistent with a different triggering mechanism for these objects (e.g. direct accretion of hot gas). In addition, characterisation of their large-scale environments found the LERGs/WLRGs are primarily hosted in cluster environments \citep{Ramos_almeida_2013, Ineson_2013, Ineson_2015}, unfavourable for galaxy mergers due to the high relative galaxy velocities \citep{Popesso_Biviano_2006}, but providing massive reservoirs of hot gas that may be accreted directly or following cooling.

\subsection{FRI/FRII}
\label{fr1_fr2_discuss}

FRI sources are almost invariably been associated with LERGs/WLRGs and HERGs/SLRGs with FRIIs. However, the mapping between the optical and radio classifications is not perfect, since a significant subset of FRII sources have been associated with LERGs/WLRGs \citep[e.g.][]{Buttiglione_2010, Tadhunter_2016}. 

It has been suggested that the FRII LERG objects may represent a switch-off or low activity phase in the HERG/SLRG lifecycle, such that the nuclear NLR emission has decreased sufficiently for a LERG/WLRG classification, but the information has yet to reach the large-scale radio lobes of the source, hence resulting in an FRII radio classification \citep[e.g.][]{Buttiglione_2010, Tadhunter_2012, Tadhunter_2016, Macconi_2020}. Indeed, there are some LERG/FRII sources in the 2Jy sample that have significantly weaker radio cores than typical HERG/FRII sources \citep{Morganti_1997}. \cite{Bernhard_2022} also found evidence that the cool interstellar medium (ISM) properties of the FRII WLRGs were comparable to the SLRGs in their sample, consistent with a switch off in the AGN activity. 

From our results, we find that there is a significant difference ($\sim$3$\sigma$) between the disturbance rates of the FRII HERGs and the FRII LERGs. When compared to their matched control samples, the FRII HERGs exhibit $>$4$\sigma$ differences, in comparison to only $\sim$1$\sigma$ for the FRII LERGs and their matched controls. Signatures of gas-rich mergers are expected to last on the order of $\sim$1 Gyr \citep{Lotz_2008}, compared with the typical AGN lifetime of $\sim$10$^6$--10$^8$ years \citep{Martini_2004}. Therefore, if the FRII HERGs and FRII LERGs were from the same parent population, the host galaxy morphology should appear the same. Instead, the FRII HERGs clearly favour a more disturbed morphology in comparison to the FRII LERGs. This suggests that the FRII LERG objects cannot solely represent a 'switched-off' phase in the HERG/SLRG lifecycle of the same galaxy population. 

There is also evidence that some FRII LERGs have stronger radio cores than expected if they were in a 'switched-off' phase \citep{Buttiglione_2010}. In addition, X-ray characterisation of their environments has revealed that the FRII LERGs appear to be hosted in richer large-scale environments on average than the FRII HERGs \citep{Ineson_2015}, also consistent with a different triggering mechanism for the FRII LERGs. Therefore, the causes of the dichotomy between the FRII HERG and FRII LERGs sources remain uncertain. 

\subsection{Merger stage}

Considering the merger stage, we find that the full 3CR sample favours late-stage mergers and interactions. This also holds consistent for most of the 3CR sub-types, except for the LERG and FRI populations, where the proportions for pre- and post-coalescence interactions are similar. 

\cite{Sanders_1988} proposed that quasars and powerful radio galaxies are triggered at the peaks of gas-rich mergers and only become visible in the post-coalescence stages, once the circumnuclear dust is expelled by the powerful winds. Winds and outflows have since been observed in many such objects \citep[e.g.][]{Rose_2018, Spence_2018, Oosterloo_2019, Lamperti_2022, Holden_2024}. In addition, \cite{Wilson_Colbert_1995} suggested that the origin of radio loudness in AGN is related to a rapidly spinning, high-mass BH, formed from the coalescence of two similar mass BHs during a major merger. This again suggests that these objects should only be visible in post-coalescence systems. Whilst we do find a preference for late-stage mergers and interactions in the majority of our radio AGN sub-samples, it is notable that a significant proportion of objects in all of our sub-samples are pre-coalescence systems (see Fig.~\ref{fig:pre_post_coal_fig} and Table~\ref{tab:pre_post_coal_table}). 

When considering objects of quasar-like luminosity in our sample, we find 35$^{+13}_{-9}$ per cent are observed in pre-coalescence systems. Similar results were found for the quasar-like luminosity radio-loud sample from \cite{ramos_almeida_2011} with 44 per cent in pre-coalescence systems \citep[percentage from][]{Pierce_2023}. In addition, it is also consistent with the intermediate-redshift sample of type 2 quasars from \cite{Bessiere_2012} (47 per cent). However, it is much lower than found for the (largely radio-quiet) Type 2 quasar sample in \cite{Pierce_2023}, where  61$^{+8}_{-9}$ per cent of the disturbed objects are located in pre-coalescence systems. This shows that quasar-like activity can be triggered across all stages of a galaxy merger, in contrast with the \cite{Sanders_1988} proposal. 

Overall, the radiatively-efficient AGN (i.e. HERGs, 3CR quasars) are pre-dominantly hosted in galaxies showing signatures consistent with late-stage mergers and interactions, whereas the radiatively-inefficient AGN (i.e. LERGs, FRIs) do not seem to have a strong preference. Signatures of both pre- and post-coalescence interactions are seen in all of the 3CR sub-samples. Therefore, we confirm the results of previous studies \citep[e.g.][]{ramos_almeida_2011, Bessiere_2012, pierce_2022, Pierce_2023} that if galaxy mergers and interactions are responsible for triggering these AGN, the activity can be triggered during different stages of the interaction and not at one specific phase. 

\subsection{Observational factors affecting merger detection rate}
\label{merger_detection_rate}

Surface brightness depth, as extensively discussed in \cite{Pierce_2023}, is a key factor in the detectability of interaction signatures and explains many of the differences between the results of similar morphological studies of AGN triggering. However, other factors may have also affected our ability to detect mergers in this study, aside from the conservative nature of our classification (as discussed in Section~\ref{herg_lerg_discussion}). 

\begin{figure}
    \centering
    \includegraphics[width=\linewidth]{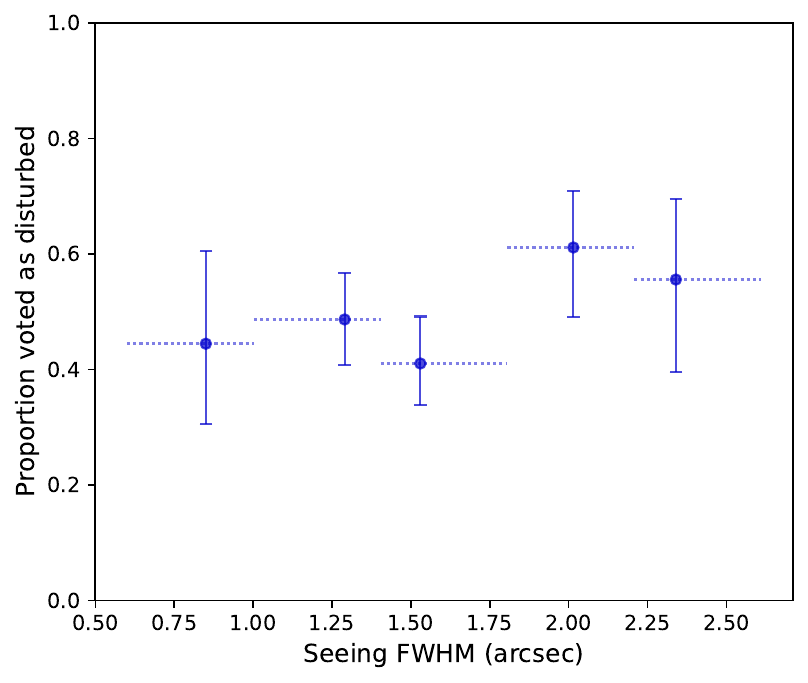}
    \caption{The proportion of the 3CR sample voted as disturbed plotted against the seeing FWHM. The proportions are plotted at the median seeing measurements for each bin.}
    \label{fig:seeing_disturb_rate}
\end{figure}

In particular, poor seeing conditions may affect the detection of faint, sharp tidal structures in the images. Fig.~\ref{fig:seeing_disturb_rate} shows the proportion of 3CR galaxies voted as disturbed against measurements of their seeing FWHM\footnote{Note that we include here the 10 3CR/2Jy objects from \cite{ramos_almeida_2011} in our study, which were observed under much better seeing conditions -- median and standard deviation of 0.80 and 0.18 arcsec, respectively -- than the rest of the objects in our sample.}. If anything, there appears to be a slight enhancement in the proportion of disturbed galaxies as the seeing becomes worse (i.e. larger seeing FWHM), although this is not statistically significant, with all values consistent within the uncertainties. Poorer seeing conditions may in fact aid the detection of certain tidal features, smoothing the structures and making it easier for the eye to detect them in the images. Therefore, we do not expect any major impacts on our results due to the inclusion of the 3CR/2Jy sources. Indeed, the disturbance rate we measure for the 3CR/2Jy objects is consistent with that we measure for the 3CR sample as a whole.

Galactic extinction due to dust may also have an impact on the detection of low surface brightness tidal features. Within the 3CR sample, 52$\pm$5 per cent of objects with an \textit{r}-band extinction of less than 1 magnitude were classified as disturbed. In contrast, for those objects with an \textit{r}-band extinction greater than 1 magnitude, only 23$^{+15}_{-8}$ per cent were classified as disturbed. Although this difference is not statistically significant, due to the smaller number of high-extinction sources and the resulting large proportion uncertainties, the results are consistent with a lower disturbance rate in the more highly extincted objects. This suggests, unsurprisingly, that a high extinction may affect our ability to detect of faint tidal features in these systems, giving a further indication that our results on the overall disturbance rates for all of the samples considered are conservative.

\section{Conclusions}
\label{conclusions}

We have conducted a deep optical imaging study using INT/WFC observations of 112 3CR galaxies with redshifts $z<0.3$ in order to investigate the dependence of the triggering mechanisms on the radio AGN sub-type. Our main results are as follows.

\begin{itemize}

    \item[(i)] The proportions of morphologically disturbed HERGs (62$^{+6}_{-7}$ per cent) and LERGs (36$^{+7}_{-6}$) show differences that are significant at a $>$3$\sigma$ level. Comparison with a stellar mass matched control sample reveals $>$4$\sigma$ differences for the HERGs. This provides strong evidence that galaxy mergers and interactions are the dominant triggering mechanism for the radiatively-efficient population of radio AGN in the local universe, reinforcing results from previous studies at a greater significance level. In contrast, the LERG results are consistent with those of their matched control sample, indicating that galaxy mergers and interactions hold much less importance for triggering the radiatively-inefficient AGN. This suggests that LERGs have a different dominant triggering mechanism, such as the accretion of gas cooling from the hot X-ray haloes of the host galaxies or galaxy clusters. 

    \item[(ii)] The FRII HERG sources are preferentially disturbed, compared with their FRII LERG counterparts -- a 3$\sigma$ difference. This provides strong evidence that the FRII LERG sources cannot simply represent a switched-off or low-activity phase of objects from the same parent galaxy population as the FRII HERGs. 
    
    \item[(iii)] We confirm the previous finding that AGN activity can be triggered during many different phases of a galaxy merger or interaction, and not at one specific phase, although a preference for late-stage mergers and interactions is found across most of the radio AGN sub-samples. 
    
\end{itemize}

Finally, we stress that this work has concentrated on the highest radio luminosity AGN, as encompassed by the 3CR sample. However, with facilities such as LOFAR, large samples of lower radio luminosity sources are now being detected. Although some work on such sources has already been done \citep[e.g.][]{Pierce_2019,pierce_2022, Gordon_2019}, it would be interesting to extend the optical morphological analysis to new emerging sub-populations of radio AGN being found in the recent surveys, such as the low radio luminosity FRII sources \citep{Mingo_2019}. This could provide important clues to how they are triggered, and their relationship to the higher luminosity radio sources.

\section*{Acknowledgements}

FB acknowledges support from the UK Science and Technology Facilities Council (STFC) [ST/Y509541/1]. JCSP acknowledges support from the UK STFC [ST/Y001249/1]. AEW acknowledges support from the UK STFC [ST/Y001257/1 and ST/X001318/1]. LRH acknowledges support from the UK STFC [ST/Y001028/1]. CRA acknowledges support from the Agencia Estatal de Investigaci\'on of the Ministerio de Ciencia, Innovaci\'on y Universidades (MCIU/AEI) under the grant ``Tracking active galactic nuclei feedback from parsec to kiloparsec scales'', with reference PID2022$-$141105NB$-$I00 and the European Regional Development Fund (ERDF). JR acknowledges financial support from the Spanish Ministry of Science and Innovation through the project PID2022-138896NB-C55 and financial support from Plan Propio de Investigación 2025 submodalidad 2.3 of the University of Cordoba. The Isaac Newton Telescope is operated on the island of La Palma by the Isaac Newton Group of Telescopes in the Spanish Observatorio del Roque de los Muchachos of the Instituto de Astrofísica de Canarias. Based on observations obtained at the international Gemini Observatory, a program of NSF NOIRLab, which is managed by the Association of Universities for Research in Astronomy (AURA) under a cooperative agreement with the U.S. National Science Foundation on behalf of the Gemini Observatory partnership: the U.S. National Science Foundation (United States), National Research Council (Canada), Agencia Nacional de Investigaci\'{o}n y Desarrollo (Chile), Ministerio de Ciencia, Tecnolog\'{i}a e Innovaci\'{o}n (Argentina), Minist\'{e}rio da Ci\^{e}ncia, Tecnologia, Inova\c{c}\~{o}es e Comunica\c{c}\~{o}es (Brazil), and Korea Astronomy and Space Science Institute (Republic of Korea). 

\section*{Data Availability}

The images of the low redshift 3CR sample used for the morphological classification in the online interface are available in the appendix. The images for the higher redshift portion of the sample and the 10 3CR/2Jy objects are available in the supplementary material of \citetalias{pierce_2022} and \cite{ramos_almeida_2011}, respectively. The full morphological classification results for the whole sample from the online interface are also presented in the appendix. Upon reasonable request to the corresponding author, further data can be made available. 



\bibliographystyle{mnras}
\bibliography{references} 


\newpage
\appendix

\section{FRI/FRII distributions}
\label{fr1_fr2_dists}

\begin{figure}
    \centering
    \includegraphics[width=\linewidth]{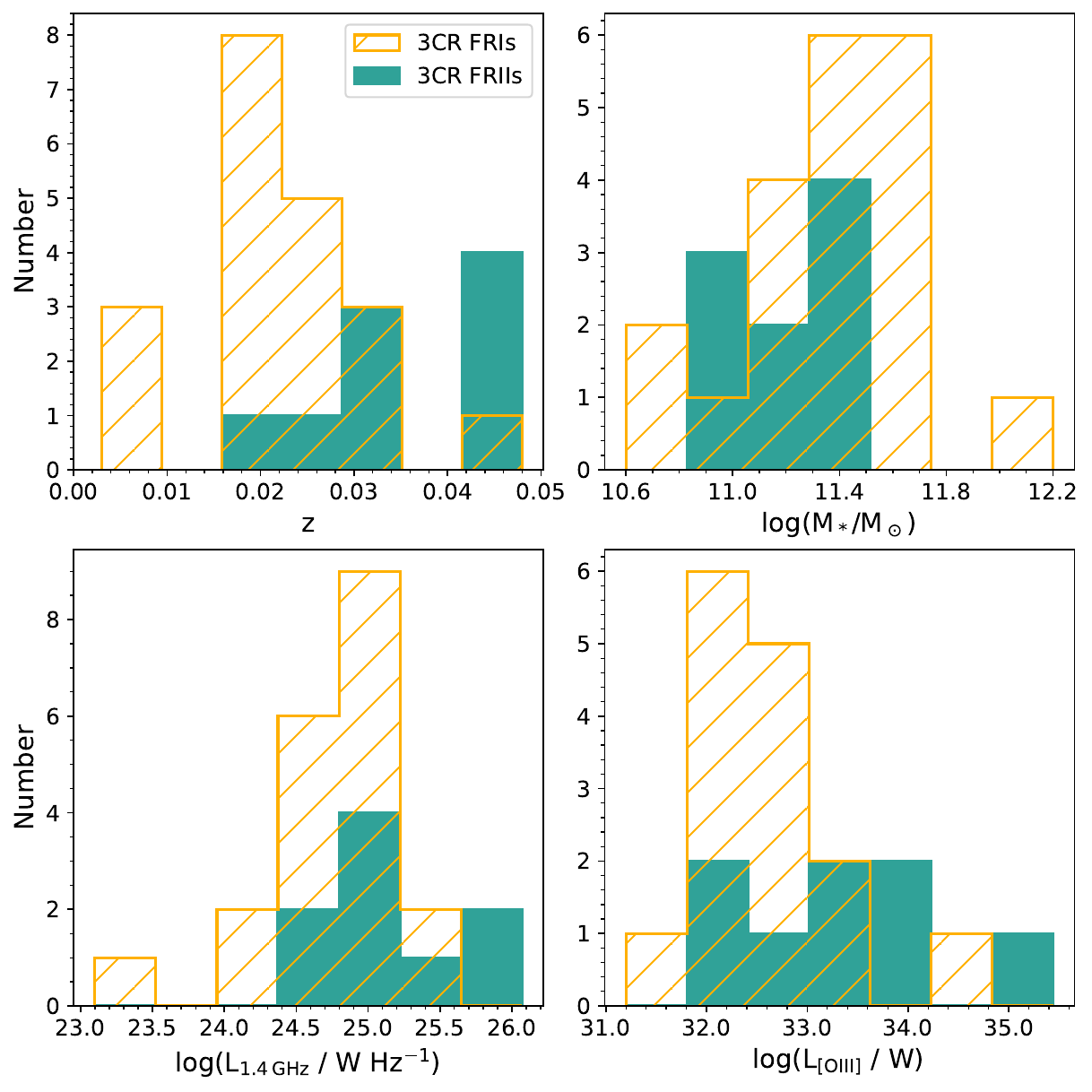}
    \caption{Distributions of redshift, stellar mass, 1.4 GHz radio luminosity and [OIII]$\lambda$5007 emission-line luminosity for the FRIs (yellow) and the FRIIs (green) in the low-redshift ($z<0.05$) 3CR sample. Sources with no stellar mass estimates were not included in the corresponding plot. Those which only had upper limits on their [OIII]$\lambda$5007 emission-line luminosity were also not considered in the distribution.}
    \label{fig:low_z_FRI_FRII_dist}
\end{figure}

\begin{figure}
    \centering
    \includegraphics[width=\linewidth]{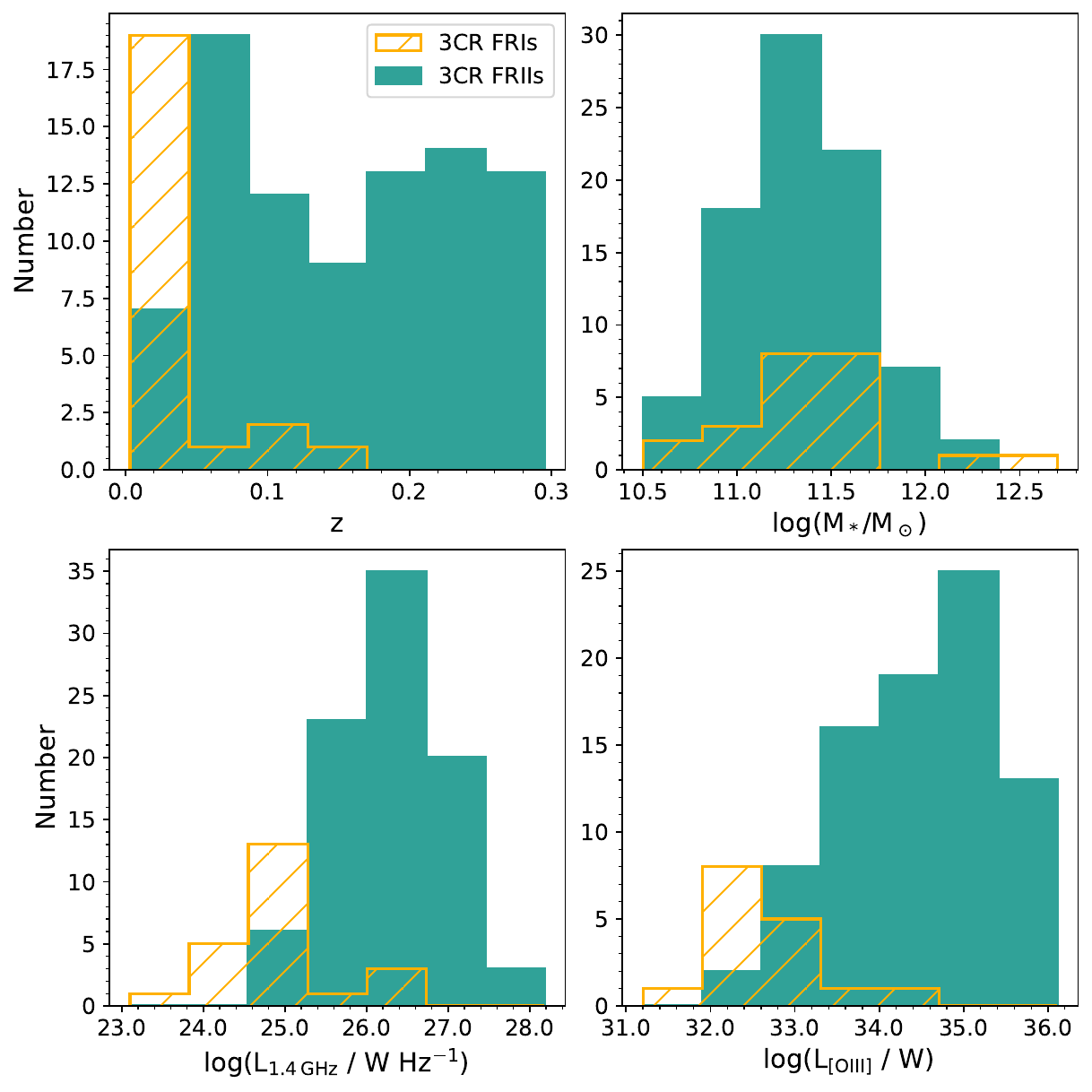}
    \caption{As in Figure~\ref{fig:low_z_FRI_FRII_dist}, but for the full 3CR sample ($z<0.3$) considered in this work.}
    \label{fig:full_FRI_FRII_dist}
\end{figure}

Redshift, stellar mass, 1.4 GHz radio luminosity and [OIII]$\lambda$5007 emission-line luminosity distributions for the FRI and FRII radio classifications in the low-\textit{z} and full 3CR sample are presented in Fig.~\ref{fig:low_z_FRI_FRII_dist} and Fig.~\ref{fig:full_FRI_FRII_dist}, respectively. 

\section{Online interface classification results}
\label{classification_results}

Table~\ref{tab:zooniverse_results} presents the detailed morphological classification results for the full 3CR sample obtained from the three questions asked to each of the classifiers, regarding: (1) whether the galaxies were disturbed or not; (2) for the disturbed galaxies, what interaction signatures were present; and (3) their morphological types (see Section~\ref{zooniverse} for a detailed explanation). We also give the optical and radio classes for the whole sample. The optical classifications for the higher redshift portion of the sample are also available in \citetalias{pierce_2022}, but we present them here for completeness, however, the radio classifications are not available so we present them here. Our radio classifications were mainly obtained from \cite{Buttiglione_2010, Buttiglione_2011}, however, these were incomplete. We have completed and updated the radio classifications here using radio maps available in the literature \citep{Birkinshaw_1985, Giovannini_1987, Wrobel_1990, Pedlar_1990, Leahy_1991, Cox_1991, Comins_1991, Black_1992, Hardcastle_1997, Ludke_1998, Gizani_2003, Venturi_2004, Dulwich_2007, Giacintucci_2007, Massaro_2012}.

\begin{table*}
	\centering
	\caption{A summary of the morphological classification results from the online interface for the full 3CR sample and their respective optical and radio classes. Column key: (1) 3CR name; (2) optical classification (HERG/LERG), with quasar-like AGN indicated (L$_{\mathrm{[OIII]}}\geq 10^{35}$ W); (3) Fanaroff-Riley radio classification; (4)-(6) final classifications determined from the three questions asked to the classifiers in the online interface (see Section~\ref{zooniverse}). Interaction signatures without brackets indicates secure morphological classifications (i.e. met a majority vote threshold), and those with brackets did not meet this threshold.}
    \label{tab:zooniverse_results}
            \begin{tabular}{lccccc} %
		\hline
            Name & Optical class & Radio class & Q1 - Disturbance & Q2 - Interaction signatures & Q3 - Host type \\
		\hline

            3C15 & LERG & 1/2 & Not disturbed & - & Elliptical \\
            3C17 & HERG & 2 & Disturbed & TIC, (A), (B), (F), (T) & Elliptical\\
            3C18 & HERG/Q & 2 & Not disturbed & - & Elliptical\\
            3C20 & HERG & 2 & Not disturbed & - & Elliptical\\
            3C28 & LERG & 2 & Not disturbed & - & Elliptical\\
            3C29 & LERG & 1 & Not disturbed & - & Elliptical \\
            3C31 & LERG & 1 & Disturbed & (A), (F), (MN), (S), (T), (TIC) & Elliptical \\
            3C33 & HERG/Q & 2 & Disturbed & A, S, (F), (T) & Elliptical \\
            3C33.1 & HERG/Q & 2 & Disturbed & MN, (B) & Elliptical \\
            3C35 & LERG & 2 & Not disturbed & - & Elliptical\\
            3C40 & LERG & 1 & Disturbed & B, MN, TIC, (A) & Elliptical \\
            3C52 & LERG & 2 & Not disturbed & - & Elliptical \\
            3C61.1 & HERG/Q & 2 & Not disturbed & - & Elliptical \\
            3C63 & HERG & 2 & Not disturbed & - & Elliptical \\
            3C66B & LERG & 1 & Disturbed & MN, (A), (S), (T), (TIC) & Elliptical \\
            3C75N & LERG & 1 & Disturbed & MN, (B) & Elliptical \\
            3C76.1  & LERG & 1 & Not disturbed & - & Elliptical \\
            3C78 & LERG & 1 & Disturbed & S, (TIC) & Elliptical \\
            3C79 & HERG/Q & 2 & Disturbed & T, (TIC) & Elliptical \\
            3C83.1 & LERG & 1 & Not disturbed & - & Elliptical \\
            3C84 & LERG & 1 & Disturbed & A, T, (D), (I), (S), (TIC) & Elliptical \\
            3C88 & LERG & 2 & Not disturbed & - & Elliptical  \\
            3C89 & LERG & 1 & Disturbed & T, (A), (F) & Elliptical  \\
            3C93.1 & HERG/Q & CSS & Disturbed & T, (A), (F) & Elliptical \\
            3C98 & HERG & 2 & Disturbed & S, (A) & Elliptical \\
            3C105 & HERG & 2 & Not disturbed & - & Elliptical \\
            3C111 & HERG/Q & 2 & Not disturbed & - & Elliptical \\
            3C123 & LERG & 2 & Not disturbed & - & Elliptical \\
            3C129 & LERG & 1 & Not disturbed & - & Elliptical \\
            3C129.1 & LERG & 1 & Not disturbed & - & Elliptical \\
            3C130 & LERG & 1 & Not disturbed & - & Elliptical \\
            3C132 & LERG & 2 & Disturbed & T, (A), (F), (I) & Elliptical \\
            3C133 & HERG/Q & 2 & Not disturbed & - & Elliptical \\
            3C135 & HERG/Q & 2 & Disturbed & MN, (A), (B), (S), (TIC) & Elliptical \\
            3C136.1 & HERG & 2 & Disturbed & I, T, (A), (D), (F) & Uncertain \\
            3C153 & LERG & 2 & Not disturbed & - & Elliptical \\
            3C165 & LERG & 2 & Not disturbed & - & Elliptical \\
            3C166 & LERG & 2 & Not disturbed & - & Elliptical \\
            3C171 & HERG/Q & 2 & Disturbed & F, T, (A), (I) & Elliptical \\
            3C173.1 & LERG & 2 & Disturbed & A, T, (F), (S), (TIC) & Elliptical \\
            3C180 & HERG/Q & 2 & Disturbed & F, (A), (I), (S), (TIC) & Elliptical \\
            3C184.1 & HERG/Q & 2 & Disturbed & T, (F), (TIC) & Elliptical \\
            3C192 & HERG & 2 & Not disturbed & - & Elliptical \\
            3C196.1 & HERG & 2 & Not disturbed & - & Elliptical \\
            3C197.1 & HERG & 2 & Not disturbed & - & Elliptical \\
            3C198 & SF & 2 & Not disturbed & - & Elliptical \\
            
	\hline
	\end{tabular}
\end{table*}

\begin{table*}
	\centering
	\caption{Continuation of Table~\ref{tab:zooniverse_results}.}
    \label{tab:zooniverse_results_cont1}
            \begin{tabular}{lccccc} %
		\hline
            Name & Optical class & Radio class &  Q1 - Disturbance & Q2 - Interaction signatures & Q3 - Host type \\
		\hline
            3C213.1 & LERG & 2 & Disturbed & A, (F), (I), (T), (TIC) & Elliptical \\
            3C219 & HERG & 2 & Disturbed & TIC, (A), (B) & Elliptical \\
            3C223 & HERG/Q & 2 & Disturbed & B, T, (A), (F), (TIC) & Elliptical \\
            3C223.1 & HERG & 2 & Disturbed & T, (A), (B), (D), (F), (TIC) & Spiral/disk \\
            3C227 & HERG & 2 & Not disturbed & - & Elliptical \\
            3C234 & HERG/Q & 2 & Disturbed & (F), (T), (TIC) & Elliptical  \\
            3C236 & LERG & 2 & Disturbed & S, (A), (F), (T) & Elliptical\\
            3C258 & HERG & CSS & Uncertain & - & Elliptical \\
            3C264 & LERG & 1 & Not disturbed & - & Elliptical \\
            3C270 & LERG & 1 & Not disturbed & - & Elliptical \\
            3C272.1 & LERG & 1 & Not disturbed & - & Elliptical \\
            3C274 & LERG & 1 & Not disturbed & - & Elliptical\\
            3C277.3 & HERG & 2 & Disturbed & S, (I), (T) & Elliptical \\
            3C284 & HERG & 2 & Disturbed & T, (A) & Elliptical \\
            3C285 & HERG & 2 & Disturbed & I, T, (A), (D), (F), (TIC) & Uncertain\\
            3C287.1 & HERG & 2 & Not disturbed & - & Elliptical\\
            3C288 & LERG & 2 & Not disturbed & - & Elliptical\\
            3C293 & LERG & 1/2 & Disturbed & D, I, T, (A), (B), (F), (MN), (S), (TIC) & Merger \\
            3C296 & LERG & 1 & Not disturbed & - & Elliptical \\
            3C300 & HERG & 2 & Disturbed & T, (A), (F), (I), (TIC) & Elliptical \\
            3C303 & HERG & 2 & Not disturbed & - & Elliptical \\
            3C303.1 & HERG/Q & CSS & Not disturbed & - & Elliptical\\
            3C305 & HERG & CSS & Disturbed & T, (F), (I), (S) & Spiral/disk \\
            3C310 & LERG & 2 & Disturbed & TIC, (A), (B), (MN) & Elliptical \\
            3C314.1 & LERG & 2 & Not disturbed & - & Elliptical\\
            3C315 & LERG & 1 & Disturbed & (A), (B), (F), (MN), (T), (TIC) & Elliptical \\
            3C317 & LERG & core-halo & Not disturbed & - & Elliptical \\
            3C318.1 & LERG & 2 & Not disturbed & - & Elliptical \\
            3C319 & LERG & 2 & Not disturbed & - & Lenticular\\
            3C321 & HERG/Q & 2 & Disturbed & MN, T, (A), (B), (F) & Merger \\
            3C323.1 & HERG/Q & 2 & Disturbed & F, (MN), (T), (TIC) & Elliptical\\
            3C326 & LERG & 2 & Disturbed & B, (A), (F), (I) & Elliptical \\
            3C327 & HERG/Q & 2 & Disturbed & S & Elliptical\\
            3C332 & HERG & 2 & Disturbed & (A), (B), (MN), (TIC) & Elliptical \\
            3C338 & LERG & 1 & Disturbed & MN, (F), (S), (TIC) & Elliptical \\
            3C346 & HERG & 1/2 & Disturbed & MN, (A), (B), (D), (F), (I), (TIC) & Elliptical \\
            3C348 & LERG & 1/2 & Not disturbed & - & Elliptical \\
            3C349 & LERG & 2 & Not disturbed & - & Elliptical\\
            3C353 & LERG & 2 & Not disturbed & - & Elliptical \\
            3C357 & LERG & 2 & Not disturbed & - & Elliptical\\
            3C371 & LERG & core-jet & Disturbed  & B, (A), (S), (TIC) & Elliptical \\
            3C379.1 & HERG & 2 & Not disturbed & - & Elliptical\\
            3C381 & HERG/Q & 2 & Disturbed & TIC, (A), (B), (MN), (T) & Elliptical \\
            3C382 & HERG & 2 & Disturbed & T, (A), (F), (I), (S), (TIC) & Elliptical \\
            3C386 & LERG & 2 & Not disturbed & - & Elliptical \\
            3C388 & LERG & 2 & Not disturbed & - & Elliptical\\
	\hline
	\end{tabular}
\end{table*}

\begin{table*}
	\centering
	\caption{Continuation of Table~\ref{tab:zooniverse_results}.}
    \label{tab:zooniverse_results_cont2}
            \begin{tabular}{lccccc} %
		\hline
            Name & Optical class & Radio class & Q1 - Disturbance & Q2 - Interaction signatures & Q3 - Host type \\
		\hline
            3C390.3 & HERG/Q & 2 & Not disturbed & - & Elliptical\\
            3C401 & LERG & 2 & Not disturbed & - & Elliptical \\
            3C402 & LERG & 1 & Not disturbed & - & Elliptical\\
            3C403 & HERG & 2 & Disturbed & S, (TIC) & Elliptical \\
            3C403.1 & LERG & 2 & Not disturbed & - & Elliptical\\
            3C405 & HERG/Q & 2 & Not disturbed & - & Elliptical\\
            3C410 & HERG/Q & 2 & Not disturbed & - & Elliptical\\
            3C424 & LERG & 1/2 & Disturbed & A, (F), (S), (TIC) & Elliptical \\
            3C430 & LERG & 2 & Not disturbed & - & Elliptical\\
            3C433 & HERG & 2 & Disturbed & A, S, TIC, (B), (I), (MN) & Merger \\
            3C436 & HERG & 2 & Disturbed & A, T, (F), (I) & Elliptical \\
            3C438 & LERG & 2 & Not disturbed & - & Elliptical\\
            3C442 & LERG & 2 & Disturbed & A, TIC, (B), (F), (MN), (S), (T) & Elliptical \\
            3C445 & HERG/Q & 2 & Disturbed & TIC, (A), (B), (F), (S) & Elliptical \\
            3C449 & LERG & 1 & Not disturbed & - & Elliptical\\
            3C456 & HERG/Q & 2 & Not disturbed & - & Elliptical \\
            3C458 & HERG/Q & 2 & Disturbed & T, (F), (TIC) & Elliptical \\
            3C459 & HERG/Q & 2 & Disturbed & A, (F), (I), (S), (T) & Elliptical \\
            3C460 & LERG & 2 & Disturbed & MN, T, (F), (I), (TIC) & Elliptical \\
            3C465 & LERG & 1 & Disturbed & MN, TIC, (A), (S), (T) & Elliptical \\
	\hline
	\end{tabular}
\end{table*}

\clearpage
\section{3CR images}
\label{stamp_images}

\begin{figure*}
    \centering
    \includegraphics[width=\linewidth]{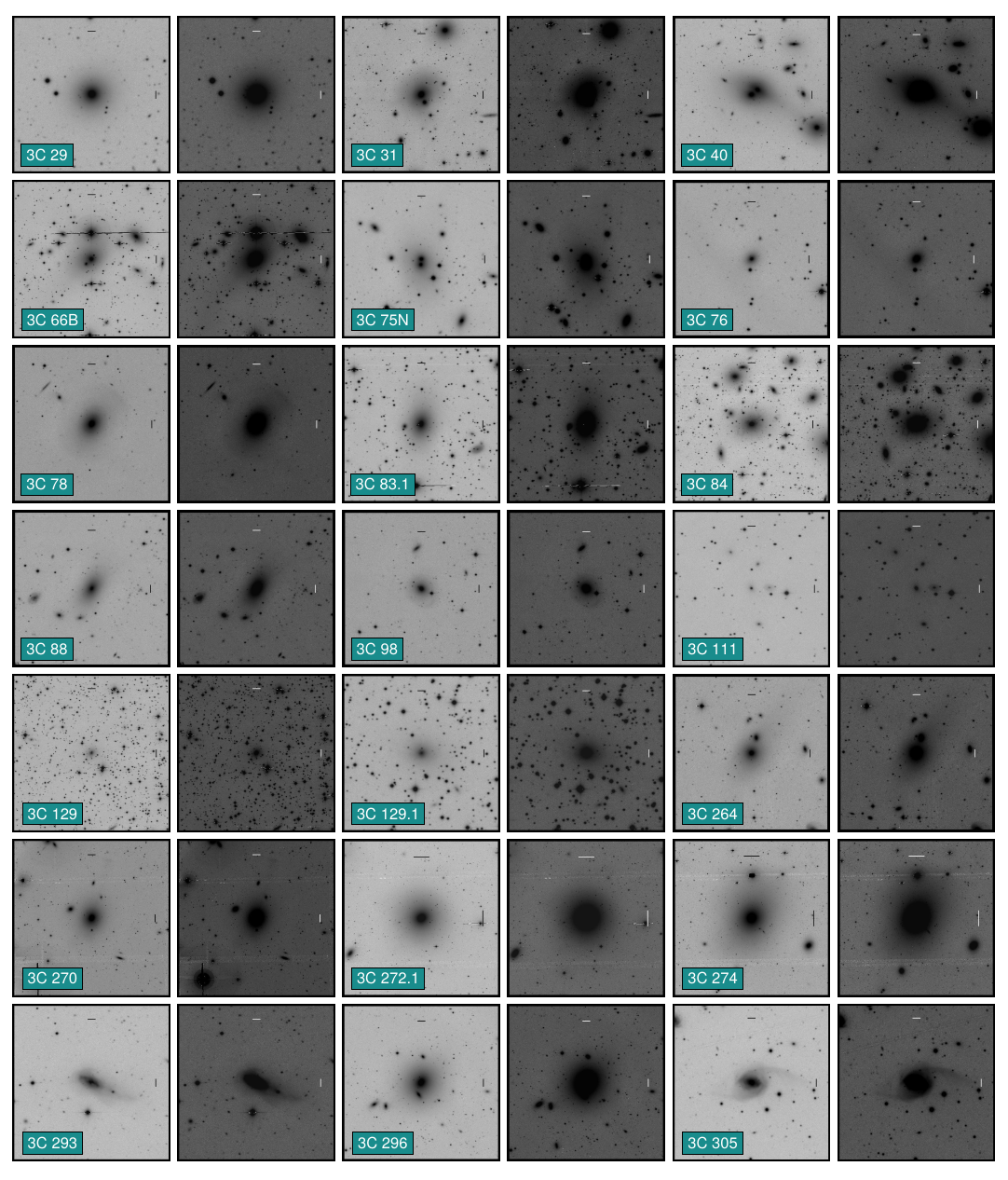}
    \caption{First set of images (200 kpc $\times$ 200 kpc) of the low-\textit{z} 3CR sample uploaded to the online interface. The high contrast (left of pair) and low contrast (right of pair) images are shown for each target.}
    \label{fig:stamps_1}
\end{figure*}

\begin{figure*}
    \centering
    \includegraphics[width=\linewidth]{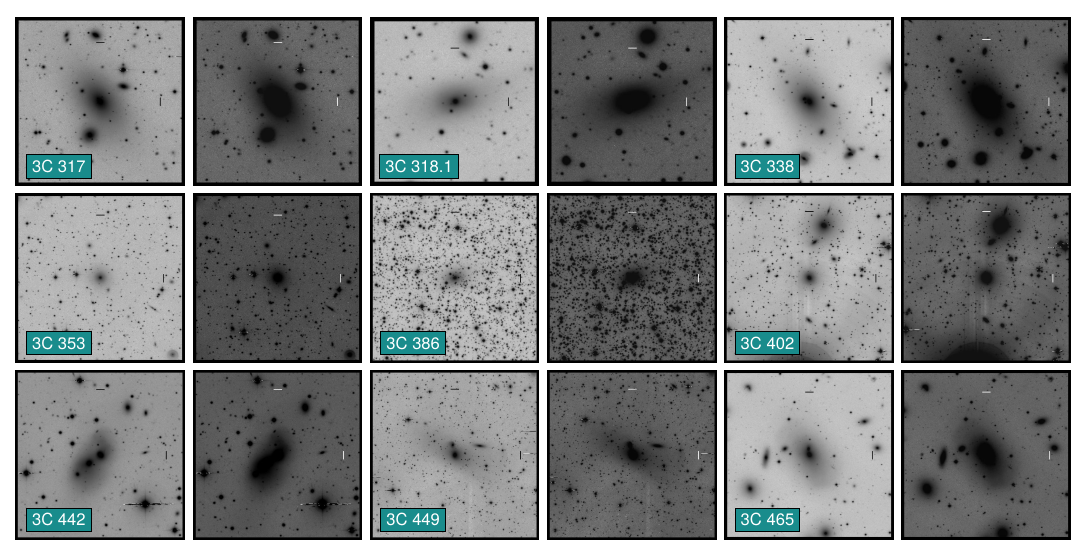}
    \caption{Second set of images (200 kpc $\times$ 200 kpc) of the low-\textit{z} 3CR sample uploaded to the online interface. The high contrast (left of pair) and low contrast (right of pair) images are shown for each target.}
    \label{fig:stamps_2}
\end{figure*}

The classification of the optical morphologies of the 112 3CR galaxies and 307 matched control galaxies, was performed using an online interface made through the Zooniverse website, as described in Section \ref{zooniverse}. Figures \ref{fig:stamps_1} and \ref{fig:stamps_2} display the high and low contrast images of the 30 low redshift ($z < 0.05$) 3CR galaxies. The images of the 72 3CRs in the sample between redshifts $0.05 < z < 0.3$, are presented in the supplementary material in \citetalias{pierce_2022}. The Gemini/GMOS-S images of the 10 3CR/2Jy objects included in our sample are available in \cite{ramos_almeida_2011}. 

\section{Proportions - SLRG/WLRG classification}
\label{SLRG_WLRG}

\begin{figure*}
    \centering
    \includegraphics[width=\linewidth]{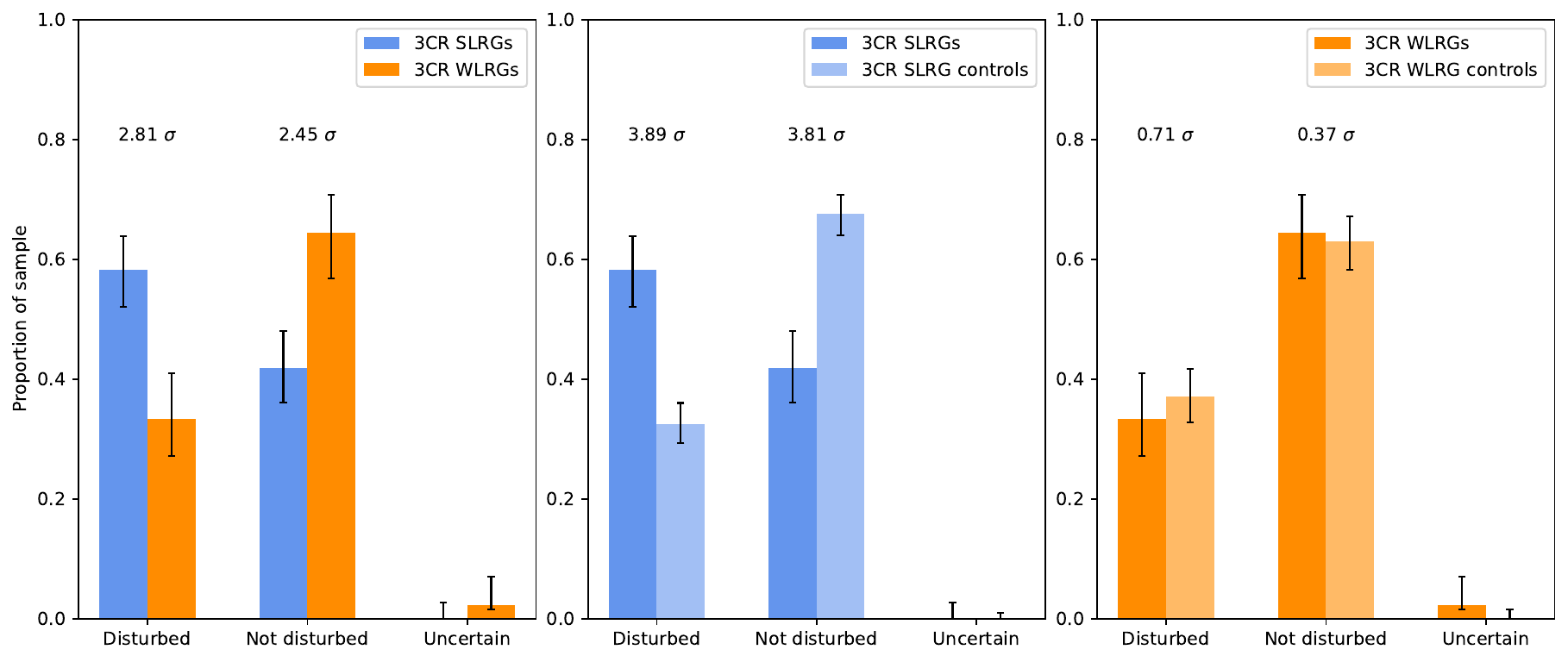}
    \caption{Proportions of the 3CR SLRGs and WLRGs classified as disturbed, not disturbed, or uncertain. The SLRGs and WLRGs are presented alongside each other (first panel), and with their respective matched control samples (second and third panel). }
    \label{fig:props_SLRG_WLRG}
\end{figure*}

Figure~\ref{fig:props_SLRG_WLRG} presents the proportions of disturbed, not disturbed and uncertain classifications for the alternative SLRG and WLRG classification used in \cite{ramos_almeida_2011}. The percentage proportions are also presented in Table~\ref{tab:slrg_wlrg_table_props}. 

\begin{table}
\def\arraystretch{1.5}
    \centering
    \caption{The proportions of the SLRGs and WLRGs in the sample which were classified as disturbed, not disturbed, or uncertain, presented alongside their matched control samples. All the proportions are presented as percentages. The number (\textit{N}) of objects in each sample is also indicated.}
    \label{tab:slrg_wlrg_table_props}
    \begin{tabular}{cccccccc}
    \hline
        & & \multicolumn{2}{c}{Disturbed (\%)} & \multicolumn{2}{c}{\makecell{Not dist. (\%)}} & \multicolumn{2}{c}{Uncertain (\%)} \\ 
        & \textit{N} & AGN & Cont. & AGN & Cont. & AGN & Cont. \\ 
    \hline
    \makecell{SLRGs\\ } & 67 & 58 $\pm$ 6 & 32 $^{+4}_{-3}$ & 42 $\pm$ 6 & 68 $^{+3}_{-4}$ & 0 $^{+3}$ & 0 $^{+1}$ \\
    
    \makecell{WLRGs\\ } & 45 & 33 $^{+8}_{-6}$ & 37 $^{+5}_{-4}$ & 64 $^{+6}_{-8}$ & 63 $^{+4}_{-5}$ & 2 $^{+5}_{-1}$ & 0 $^{+2}$ \\
    \hline
    \end{tabular}
\end{table}


\bsp	
\label{lastpage}
\end{document}